\documentclass[a4paper,11pt]{article}
\usepackage{booktabs}
\usepackage{color}

\usepackage{amsthm,amsmath, amssymb, amsfonts, amscd, xspace, pifont, color, bm}
\usepackage{mathtools,etoolbox}
\usepackage{epsfig, psfrag}
\usepackage{float, fancybox, fullpage}
\usepackage{delarray}
\usepackage{hyperref}
\usepackage{url}
\usepackage{anysize}
\usepackage{multirow}
\usepackage{geometry}
\usepackage{rotating}
\usepackage{graphicx}
\usepackage{caption}
\usepackage{subcaption}
\usepackage[round]{natbib} 
\usepackage[T1]{fontenc}
\usepackage[utf8]{inputenc}
\usepackage{authblk}
\usepackage{xcolor}
\usepackage{algorithm}
\usepackage{algpseudocode}
\usepackage{pifont}
\usepackage[algo2e]{algorithm2e} 
\usepackage{enumitem}
\usepackage[short]{optidef}
\usepackage{physics}
\usepackage{relsize}
\usepackage[export]{adjustbox}
\usepackage{caption}
\usepackage{subcaption}
\usepackage{xparse}
\usepackage{dsfont}
\usepackage{siunitx}

\newcommand{\ze}{Z\kern-0.45emZ}
\newcommand{\esp}{I\kern-0.37emE}
\newcommand{\N}{I\kern-0.37emN}
\newcommand{\one}{ {\rm 1\kern-0.19eml} }
\newcommand{\realset}{I\kern-0.37emR}

\newcommand{\bx}{\boldsymbol{x}}

\newcommand{\bX}{\boldsymbol{X}}

\newcommand{\bv}{\boldsymbol{v}}

\newcommand{\bmu}{\mbox{\boldmath $\mu$}}

\SetCommentSty{mycommfont}

\makeatletter
\newcommand{\ssymbol}[1]{^{\@fnsymbol{#1}}}
\makeatother

\NewDocumentCommand{\INTERVALINNARDS}{ m m }{
    #1 {,} #2
}
\NewDocumentCommand{\interval}{ s m >{\SplitArgument{1}{,}}m m o }{
    \IfBooleanTF{#1}{
        \left#2 \INTERVALINNARDS #3 \right#4
    }{
        \IfValueTF{#5}{
            #5{#2} \INTERVALINNARDS #3 #5{#4}
        }{
            #2 \INTERVALINNARDS #3 #4
        }
    }
}

\SetKwInput{KwInput}{Input}                
\SetKwInput{KwOutput}{Output}              

\newtheorem{theorem}{Theorem}

\DeclarePairedDelimiterX\Card[1]\lvert\rvert{
  \ifblank{#1}{{-}}{#1}
}

\DeclarePairedDelimiter\floor{\lfloor}{\rfloor}

\newcommand{\norms}[1]{\left\lVert#1\right\rVert}

\def\blue{\textcolor{black}}
\def\orange{\textcolor{black}}

\begin{document}
\title{Linear Algorithms for Robust and Scalable Nonparametric Multiclass Probability Estimation}
\date{}

\author[1,2]{Liyun Zeng}
\author[1,2,*]{Hao Helen Zhang}
\affil[1]{Statistics and Data Science GIDP, University of Arizona, Tucson, AZ, 85721, USA}
\affil[2]{Department of Mathematics, University of Arizona, Tucson, AZ, 85721, USA}
\affil[*]{Corresponding author: hzhang@math.arizona.edu}

\maketitle

\begin{abstract}
Multiclass probability estimation is the problem of estimating conditional probabilities of a data point belonging to a class given its covariate information. It has broad applications in statistical analysis and data science. Recently a class of weighted Support Vector Machines (wSVMs) has been developed to estimate class probabilities through ensemble learning for $K$-class problems \citep*{WuZhaLiu2010, wang_multiclass_2019}, where $K$ is the number of classes. The estimators are robust and achieve high accuracy for probability estimation, but their learning is implemented through pairwise coupling, which demands polynomial time in $K$. In this paper, we propose two new learning schemes, the baseline learning and the One-vs-All (OVA) learning, to further improve wSVMs in terms of computational efficiency and estimation accuracy. In particular, the baseline learning has optimal computational complexity in the sense that it is linear in $K$. Though not the most efficient in computation, the OVA is found to have the best estimation accuracy among all the procedures under comparison.  The resulting estimators are distribution-free and shown to be consistent. We further conduct extensive numerical experiments to demonstrate their finite sample performance.
\end{abstract}

\vspace{0.15in} \noindent {\em Key Words and Phrases:} support vector machines, multiclass classification, probability estimation, linear time algorithm, non-parametric, scalablility.

\section{Introduction}

In machine learning and pattern recognition, the goal of multiclass classification is to assign a data point to one of $K$ classes based on its input, where $K \ge 3$. In specific, we observe the pair $(\bX, Y)$ from an unknown distribution \orange{$P(\bx, y)$}, where $\bX\in\realset^p$ is the input vector and $Y\in\mathcal{A}=\left\{1,2,\ldots,K\right\}$ indicates class membership, and
the task is to learn a discriminant rule $\phi:\realset^p\rightarrow\mathcal{A}$ for making predictions on future data. Define the conditional class probabilities as \orange{$p_j(\bx)=P(y=j|\bX=\bx)$} for $j \in \{1,\ldots,K\}.$ For any classifier $\phi$, its prediction performance can be evaluated by its risk $R(\phi)=\mathop\mathbb{E}_{(\bx,y)\sim P}[I(y\neq\phi(\bx))]=P(y\ne\phi(\bx))$, also called the expected prediction error (EPE).
The optimal rule minimizing the risk is the \emph{Bayes} classifier, given by:
\begin{equation}
\phi_B(\bx)=\arg\max_{j=1,\ldots,K} p_j(\bx).
\end{equation}

For classification purposes, although it is sufficient to learn the $\mbox{argmax}$ rule only, the estimation of class probabilities is usually desired as they can quantify uncertainty about the classification decision and provide additional confidence. Probability estimation is useful in a broad range of applications. For example, in medical decision making such as cancer diagnosis, class probabilities offer an informative measurement of the reliability of the disease classification \citep{Hastie2009}. In e-commerce advertising and recommendation systems, accurate estimation of class probabilities enables businesses to focus on the most promising alternatives and therefore optimize budget to maximize profit gain and minimize loss. Moreover, for unbalanced problems, class probabilities can help users construct optimal decision rules for cost-sensitive classification, where different types of misclassification errors may incur different costs \citep{Herbei2006}.

In literature, there are a variety of methods for probability estimation, and commonly used methods include linear discriminant analysis (LDA), quadratic discriminant analysis (QDA), and logistic regression. These methods typically make certain assumptions on statistic models or data distributions \citep{McCullagh89}. 
Recently, multiclass SVMs \citep{Vapnik98, WestionWatkins99, Crammer2001OnTA, Lee2004, WangShen2007, Liu2007, LiuYuan2011, ZhangLiu2013, Huang2013, lei2015multiclass} have shown great advantage in many real-world applications, such as cancer diagnosis, hypothyroid detection, handwritten digit recognition, spam detection, \orange{speech recognition}, and online news classification \citep{burges1998, cristi00, Zhu04, 6046926,Freha4481, saigal_2020}. However, standard SVMs can not estimate class probabilities. To tackle this limitation, the concept of binary weighted SVMs (wSVMs) is proposed by \cite*{WSL2008} to learn class probabilities via ensemble learning. In particular, the binary wSVM methods assign different weights to data from two classes, train a series of classifiers from the weighted loss function, and aggregate multiple decision rules to construct the class probabilities. The wSVM methods are model-free, hence being both flexible and robust.
\cite*{WuZhaLiu2010} extend the wSVM from $K=2$ to $K\ge 3$ by learning multiple probability functions simultaneously, which has nice theoretical properties and empirical performance but demands computational time exponentially in $K$, making it impracticable in \orange{many real-world} applications with $K \ge 4$. 
Recently, \cite*{wang_multiclass_2019} propose the divide-and-conquer learning approach via pairwise coupling to reduce the \orange{computational} cost of \cite*{WuZhaLiu2010} from exponential to polynomial \orange{time} in $K$, but it is still not optimal for complex problems with a large $K$, say, $K\ge 5$.  

In this article, 
we \blue{propose new estimation schemes and computational algorithms to further improve \cite*{wang_multiclass_2019} in terms of its computation and estimation accuracy. The first algorithm uses the idea of ``baseline learning'', which can reduce the computational time of wSVMs from polynomial to linear \orange{complexity} in $K$, and \orange{it} is therefore scalable for learning massive data sets with a large \orange{number of classes} $K$ and \orange{sample points} $n$. The second scheme employs the One-vs-All strategy to improve wSVMs in terms of probability estimation accuracy}. Both methods can be formulated as Quadratic Programming (QP), and solved by popular optimization packages in R, Python, MATLAB, and C++ \emph{etc}. \blue{We also provide rigorous analysis on their computational complexity}.  In addition, \blue{Their divide-and-conquer nature can further speed up their implementation} via parallel computing by GPU and multi-core parallel computing, high-performance computing (HPC) clusters, and massively parallel computing (MPP). 

The rest of the paper is organized as follows. \blue{Section \ref{sec:review} introduces the notations reviews the multiclass wSVMs framework. Section \ref{sec:3.1} proposes the baseline learning scheme and shows its computational and theoretical properties. Section \ref{sec:4} introduces the One-vs-All learning scheme and presents the computational algorithm and Fisher consistency results.} Section \ref{sec:algcomplx} conducts a rigorous analysis of the computational complexity of the new learning schemes and algorithms. We show that the baseline learning method enjoys the optimal complexity, with \orange{computational} time linear in the number of classes $K$.
Section \ref{sec:tuning} discusses the issue of hyperparameter tuning.  Sections \ref{sec:simu} and \ref{sec:real-w} \blue{are contributed to numerical studies by performing extensive simulated} and real-world data analysis and comparison. Section \ref{sec:cond} presents concluding remarks and discussions. 

\section{Notations and Review of Weighted SVMs} \label{sec:review}
\subsection{Notations}
Denote the data points by $\left\{\left(\bx_i,y_i\right), i=1,\ldots,n\right\}$, where $\bx_i=(x_{i1}, \ldots, x_{ip})^{\small\mathrm{T}}\in\realset^p$, $n$ is the sample size, and $p$ is the dimension of input $bx$. There are $K$ classes, with $K\ge 3$. For $j=1,\ldots, K$, denote the subset of data points belonging to class $j$ by
$\mathcal{S}_j = \{(\bx_i, y_i) \mid y_i = j; ~~ i=1, \ldots, n\}$, and the corresponding sample size by $n_j=\Card{\mathcal{S}_j}$. Denote the multiclass learner trained from the data by $\bm{\hat{f}}= \{\hat{f}_1,\ldots,\hat{f}_K\}$, where $\hat{f}_j(\bx)$ represents evidence of a data point with input $\bx$ belonging to class $j$. The classifier is constructed using the argmax rule, i.e., $\hat{\phi}(\bx)= \arg\max_{1 \le j \le K}\hat{f}_j(\bx)$.

\subsection{Binary Weighted SVMs}
\label{ssec:1}

In binary class problems, the class label $y\in\{+1, -1\}$.  Define the conditional class probabilities $p_{+1}(\bx)=P(Y=+1|\bX=\bx)$ and $p_{-1}(\bx)=P(Y=-1|\bX=\bx)$ for class $+1$ and $-1$, respectively. The standard SVM learns the decision function $f\in\mathcal{F}$ by solving the regularization problem:
        \begin{equation}
        \min_{f\in\mathcal{F}}\enskip\frac{1}{n}\sum_{i=1}^nL(y_if(\bx_i))+\lambda J(f), \label{svm}
        \end{equation}
where $L(yf(\bx))=[1-yf(\bx)]_+=\max\{0, 1-yf(\bx)\}$ is the hinge loss, $\mathcal{F}$ is some functional space, the regularization term $J(f)$ controls model complexity, and the tuning parameter $\lambda>0$ balances the bias–variance tradeoff \citep{Hastie2009}. For linear SVMs, $f(\bx)=\beta_0 +\boldsymbol{\beta}^{\small\mathrm{T}}\bx$. For kernel SVMs, $f$ employs a bivariate kernel representation form $f(\bx)=d +\sum_{i=1}^n c_i\mathbf{K}(\bx_i, \bx)$, according to the representer theorem \citep{KW1971}, where $\mathbf{K}(\cdot, \cdot)$ is a Mercer kernel, $\mathcal{F}=\mathcal{H}_\mathbf{K}$ is the reproducing kernel Hilbert space \citep[RKHS,][]{Wahba90} induced by $\mathbf{K}(\cdot,\cdot)$, and $J(f)=\norms{{\bm{f}}}_{\mathcal{H}_\mathbf{K}}^2=\sum_{i=1}^n\sum_{j=1}^{n}c_ic_j\mathbf{K}(\bx_i, \bx_j)$. 
Then the optimization problem \eqref{svm} amounts to
        \begin{equation}
            \min_{d, c_1,\cdots, c_n}\enskip\frac{1}{n}\sum_{i=1}^nL(y_if(\bx_i))+\lambda
            \sum_{i=1}^{n}\sum_{j=1}^{n}c_ic_j\mathbf{K}(\bx_i, \bx_j),  \quad \mbox{where} ~
            f(\bx)=d+\sum_{i=1}^nc_i \mathbf{K}(\bx_i, \bx).
            \label{svm2}
        \end{equation}
It is known that the minimizer of the expected hinge loss $\mathop\mathbb{E}_{(\bX,Y) \sim \orange{P(\bX, Y)}} [1-Yf(\bX)]_+$ has the same sign as the Bayes decision boundary $\mbox{sign}[p_{+1}(\bx)-\frac{1}{2}]$
\citep*{lin02}, therefore binary
SVMs directly target on the optimal Bayes classifier without estimating $p_{+1}(\bx)$.

Despite their success in many real-world classification problems, standard SVMs can not estimate the conditional class probabilities $p_{+1}(\bx)$ and $p_{-1}(\bx)$. To overcome this limitation, binary weighted SVMs \citep[wSVMs,][]{WSL2008} were proposed to learn a series of classifiers by minimizing the weighted hinge loss function, and construct probabilities by aggregating multiple decision rules. In particular, by assigning the weight $\pi$ $(0\leq\pi\leq1)$ to data points from class $-1$ and assigning $1-\pi$ to those from class $+1$, the wSVM solves the following optimization problem
        \begin{equation}
          \min_{f\in\mathcal{H}_\mathbf{K}}\enskip\frac{1}{n}\Big[(1-\pi)\sum_{y_i=1}L({y_if(\bx_i)})+\pi\sum_{y_i=-1}L({y_if(\bx_i)})\Big]+\lambda J(f).
            \label{wsvm1}
        \end{equation}
 For any $\pi$, the minimizer of the expected weighted hinge loss $\mathop{\mathbb{E}}_{(\bX,Y)\sim \orange{P(\bX, Y)}}\left\{W(Y)L[Yf(\bX)]\right\}$
has the same sign as sign[$p_{+1}(\bX)-\pi$] \citep*{WSL2008}. To estimate the class probabilities, we first obtain multiple classifiers \orange{$\hat{f}_{\pi_2},\cdots, \hat{f}_{\pi_{M}}$} by solving \eqref{wsvm1} with a series of values \orange{$0=\pi_1<\pi_2<\cdots<\pi_{M}<\pi_{M+1}=1$.} 
Given any point $\bx$, the values
$\hat{f}_\pi(\bx)$ is non-increasing in $\pi$, so there exists a unique \orange{$m^* \in \{1, \ldots, M\}$} such that 
$\pi_{m^*}<p_{+1}(\bx)<\pi_{m^*+1}$. Consequently, we can construct a consistent probability estimator by $\hat{p}_{+1}(\bx)=\frac{1}{2}(\pi_{m^*}+\pi_{m^*+1})$. The numerical precision level of $\hat{p}_{+1}(\bx)$ is determined by $M$, which controls the fineness of the grid points $\pi$'s.

\subsection{Multiclass Weighted SVMs.} \label{ssec:2}

Assume $Y=y\in\{1, 2, \ldots, K\}$ with $K\ge 3$. The goal is to estimate class probabilities \orange{$p_j(\bx)=P(Y=j|\bX=\bx)$ for $j \in \{1,\ldots, K\}$}. \cite*{WuZhaLiu2010} extended wSVMs from $K=2$ to $K\ge 3$ via a joint estimation approach. However, its computational complexity is exponential \blue{in} $K$, which makes the procedure infeasible for problems with a large $K$ and a large $n$. Recently \cite*{wang_multiclass_2019} proposed the pairwise coupling method which decomposes the $K$-class problem into $\binom{K}{2}$ binary classification problems and enjoys a reduced computational complexity, polynomial in $K$. For each class pair $(j, k)$ with $1\le j \neq k\le K$, the pairwise coupling first learns the pairwise conditional probability $\hat{q}_{j|(j, k)}$ by training binary wSVMs, and then aggregates the decision rules from all pairs to construct the class probabilities estimator as
           \begin{equation}
                \hat p_{j}(\bx)=\frac{\hat q_{j|(j, k)}(\bx)/\hat q_{k|(j,k)}(\bx)}{\sum_{s=1}^{K} \hat q_{s|(s, k)}(\bx)/\hat q_{k|(s,k)}(\bx)}, \quad j = 1, \cdots, K; ~~ \blue{k \neq j}. \label{allprob1}
            \end{equation}
\blue{Class prediction is done by either the argmax rule}, i.e., $\hat y = \arg \max_{1 \le j \le K}\hat{p}_{j}(\bx)$, or the max voting algorithm \citep{6221261,tomar2015,ding2019}.
\cite*{wang_multiclass_2019} showed that the wSVMs can outperform benchmark methods like kernel multi-category logistic regression \citep{ZH05}, random forest, and classification trees. 

\section{New Learning Scheme: Baseline Learning}
\label{sec:3.1}

\subsection{Methodology}\label{ssec:3.1}

\blue{The pairwise coupling  method \citep{wang_multiclass_2019} reduces the computational complexity of \cite{WuZhaLiu2010} from exponential to polynomial in $K$. The procedure works well for a moderate $K$, but not feasible for a much larger $K$. To speed up, we propose a more efficient learning scheme, the baseline learning. We show that the computational cost of the baseline learning is linear in $K$, making it scalable for classification problems with a large $K$. }

\blue{In pairwise coupling, we \orange{optimize the algorithm (discussed in Section \ref{sec:simu})} to choose adaptively the \orange{best} baseline class for each data point. It requires training all class pairs $(j,k)$ with $j \neq k$ and fitting $\binom{K}{2}$ binary \orange{wSVMs} problems. To speed up, we propose using a common baseline class $k^*$ and training only $K-1$ binary problems for $(j, k^*)$ with $j\neq k^*$, call the procedure as baseline learning. The rational behind the baseline learning is that, the final \orange{class} probability estimator does not depend on the choice of the baseline class \orange{as shown in \eqref{allprob1}} \citep{wang_multiclass_2019}. In Section \ref{sec:algcomplx}, the baseline learning is shown to enjoy computational complexity linear in $K$, which is the lowest among all the wSVMs \orange{methods}.} 
After choosing $k^*$, we train $K-1$ binary wSVMs problems and obtain $K-1$ pairwise conditional probabilities $\hat{q}_{j|(j, k^*)}$, for $j\neq k^*$ and $j=1,\ldots, K$. Assuming $\hat{q}_{k^*|(k^*, k^*)}=1$, the final class probability \orange{estimators} are then given by: 
           \begin{equation}
                \hat{p}_{j}^{bs}(\bx)=\frac{\hat q_{j|(j, k^*)}(\bx)/\hat q_{k^*|(j,k^*)}(\bx)}{\sum_{s=1}^{K} \hat q_{s|(s,k^*)}(\bx)/\hat q_{k^*|(s,k^*)}(\bx)}, \quad j=1, \cdots, K.
                \label{eqn:allprobBS1}
            \end{equation}
            
The choice of $k^*$ is critical to optimize finite performance of the baseline learning. To assure numerical stability, we need to choose $k^*$ such that the denominators $\hat{q}_{k^*|(j, k^*)}$ of the ratios in \eqref{eqn:allprobBS1} are kept away from zero simultaneously for all the classes $j$ with $j \neq k^*$. Towards this, we propose the following two methods of choosing $k^*$ and then evaluate their finite sample performance in numerical studies. 

\paragraph*{Choice of $k^*$ (Method 1)}
 We propose choosing $k^*$ to be the largest subclass in the training set, i.e., $k^* = \arg \max_{1 \le j \le K} \Card{\mathcal{S}_j}$. With this choice, the quantity $\hat{q}_{k^*|(j, k^*)}$ is relatively large among all possible choices and hence stablizes the ratios in \eqref{eqn:allprobBS1}. This method has been used in other contexts of one-class and multi-class SVMs classification tasks and found computationally robust \citep{krawczyk2014}. We will refer to the baseline learning wSVMs with this choice as B1-SVM.
 
\paragraph*{Choice of $k^*$ (Method 2)} Alternatively, we propose choosing $k^*$ to be the class not too far from any other class, in order to prevent $\hat{q}_{k^*|(j, k^*)}$ in \eqref{eqn:allprobBS1} from being too close to zero for any $j\neq k^*$. Towards this, we would look for the class which sits at the ``middle'' location of all the classes. For each class $j$, we compute its within-class compactness $\mathcal{D}_{cp}$ and between-class distances $\mathcal{D}_{bc}$ \citep{islam_2016} to another class $j'$, construct an ``aggregated'' distance $\mathcal{D}_{agg}(j)$ to measure its overall distance to all the remaining classes. The class $k^*$ is chosen based on the median of $\mathcal{D}_{agg}(j)$'s.

\medskip
For each class $j$, we compute its within-class compactness $\mathcal{D}_{cp}(j)$ using $\mathcal{S}_j$ as follows: (1) for each point $\bx_t \in \mathcal{S}_j$, $1 \le t \le \Card{\mathcal{S}_j}$, calculate the sum of its Euclidean distances to all other data points in class $j$, i.e.  
$d_{t} = \sum_{\forall \bx_s \in \mathcal{S}_j\setminus\{\bx_t\} }\norms{{\bx_s - \bx_t}}_{2}$; (2) identify the median point $\tilde{\bx}_{m}$, corresponding to the median of $\{d_1, \ldots, d_{\small\Card{\mathcal{S}_j}}\}$; (3) calculate the maximum distance of the points in class $j$ to the median, i.e., 
$\text{dm}(j) = \max_{\bx_s\in\mathcal{S}_j}\norms{{\bx_s - \tilde{\bx}_{m}}}_{2}$; (4) define $\mathcal{D}_{cp}(j) = \text{dm}(j)$. 
For each pair $(j,j'), 1 \le j \neq j' \le K$, their between-class distance $\mathcal{D}_{bc}(j,j')$ is the minimum distance between the boundary points of the two classes. In specific, it is equal to the minimum value of the Euclidean distances of data points from class $j$ to those in class $j'$, i.e., $\mathcal{D}_{bc}(j,j') = \min_{{\forall \bx_s \in \small\{\mathcal{S}_j\}}\times{\forall \bx_t \in \small\{\mathcal{S}_{j'}\}}} \norms{{\bx_s - \bx_t}}_{2}$. Finally, the aggregated class distance $\mathcal{D}_{agg}(j)$ is defined as
\begin{equation}\label{agg_dis}
    \mathcal{D}_{agg}(j) = \frac{1}{K}
    \frac{\sum_{\small\{j' \mid 1 \le j' \neq j \le K\}}\mathcal{D}_{bc}(j,j') }{\mathcal{D}_{cp}(j)}.
\end{equation}

\noindent
Define $k^* = \arg\mbox{median}(\{\mathcal{D}_{agg}(j) \mid 1 \le j \le K\})$. This wSVMs learner is referred to as B2-SVM.
            
\subsection{Computational Algorithm}\label{ssec:3.2}
To implement the baseline learning, we carry out the following steps: (1) choose the common baseline class $k^*$; (2) decompose the $K$-class classification problem into $K-1$ binary problems; (3) for each $j$, train the binary wSVMs for the pair $(j, k^*)$ and obtain the pairwise conditional probability estimator $\hat{q}_{j|(j, k^*)}$; (4) compute $\hat p_{j}^{bs}(\bx)$ using \eqref{eqn:allprobBS1} for each $j\neq k^*$. The following is the detailed algorithm.
\medskip
 \begin{enumerate}[labelwidth=1.5cm,labelindent=10pt,leftmargin=1.7cm,label=\bfseries Step \arabic*.,align=left]
 \item[Step 1:] Choose the baseline class $k^*$, using \orange{either} Method 1 (B1-SVM) or Method 2 (B2-SVM).
 
 \item[Step 2:] For each $j\neq k^*$, define a function $R_{j, k^*}(y)=1$ if $y=j$; $R_{j, k^*}(y)=-1$ if $y=k^*$. Fit the kernel binary wSVMs for the class pair $(j, k^*)$ by solving the optimization problem:
            \begin{equation}\label{obj8ba}
                \min_{f\in \mathcal{H}_\mathbf{K}}\enskip \frac{1}{n}\Big[(1-\pi_m)\sum_{y_i=j}L(R_{j,k^*}(y_i)f(\bx_i))+\pi_m\sum_{y_i=k^*}L(R_{j, k^*}(y_i)f(\bx_i))\Big]+\lambda J(f)
            \end{equation}
over a a series \orange{$0 <\pi_2<\cdots<\pi_{M}<1$. For each $m=2,\ldots, M$,} denote the optimized classifier of \eqref{obj8ba} as $\hat f_{j, k^*, \pi_m}(\cdot)$. \orange{Furthermore, we define $\pi_1 = 0$, $\pi_{M+1} = 1$, and assume $\hat f_{j, k^*, \pi_m}(\cdot) >0$ if $m=1$ and $\hat f_{j, k^*, \pi_m}(\cdot) < 0$ if $m=M+1$.}
\item[Step 3:]  For each class pair $(j, k^*)$, calculate the pairwise conditional probability estimator as:
\begin{equation}\label{eqn:argprob2}
    \hat{q}_{j|(j, k^*)}(\bx)= \frac{1}{2}\left[\arg\min_{\pi_m}\{\hat f_{j, k^*, \pi_m}(\bx) < 0\}+\arg\max_{\pi_m}\{\hat f_{j, k^*, \pi_m}(\bx) > 0\}\right],  \quad  \forall\bx\in\mathcal{S}.
\end{equation}
 
\item[Step 4:]  Calculate the Bayesian posterior class probabilities estimator as:
\begin{align}
    \label{eqn:allprob2b}
    \begin{split}
     \hat p_{j}^{bs}(\bx)&=\frac{\hat q_{j|(j, k^*)}(\bx)/\hat q_{k^*|(j,k^*)}(\bx)}{\sum_{s=1}^{K} \hat q_{s|(s, k^*)}(\bx)/\hat q_{k^*|(s,k^*)}(\bx)}, \quad j=\{1, \cdots, K\} \setminus \{k^*\},
    \\
     \hat p_{k^*}^{bs}(\bx)&= 1 -  \sum_{\{j,1\le j \neq k^*\le K \}}\hat{p}_{j}^{bs}(\bx).
    \end{split}
\end{align}
\end{enumerate}

Since the baseline \orange{learning} trains only $K-1$ binary wSVMs, compared to $\binom{K}{2}$ wSVMs classifiers in the pairwise coupling method, its \orange{computational} saving is substantial for a large $K$. In oder to obtain optimal result, For now, the regularization parameter $\lambda>0$ needs to be tuned adaptively using the data at Step 2. In Section \ref{sec:tuning}, we will discuss the parameter tuning in details.

\subsection{Theory}\label{ssec:3.3}
We now study asymptotic properties of the conditional probability estimators given by the baseline learning \eqref{eqn:allprobBS1} and establish its Fisher consistency. At Step 2, since the baseline learning solves the sub-problems of the pairwise coupling, Lemma 2 in \cite*{wang_multiclass_2019} holds in our case as well. Therefore, for any fixed $j$ and $\pi$, the global minimizer of the weighted hinge loss corresponding to \eqref{obj8ba} for pair $(j,k^*)$ is a consistent estimator of $q_{j|(j, k^*)}(\bX)-\pi$. When the sample size $n \rightarrow \infty$ and the functional space $\mathcal{F}$ is sufficiently rich, with a properly chosen $\lambda$, the solution to \eqref{obj8ba}, $\hat{q}_{j|(j, k^*)}(\bx)$, will converge to $q_{j|(j,k^*)}(\bx)$ asymptotically. Theorem \ref{thm1} below shows that the estimator from the baseline learning wSVM, given by  \eqref{eqn:allprob2b}, is asymptotically consistent under certain regularity conditions. Proof of Theorem 1 is similar to those in \cite*{Wang10_2307} and \cite*{WSL2008} and hence omitted.

\begin{theorem}\label{thm1} Consider the $K$-class classification problem with kernel weighted SVMs. Denote the class probabilities estimator as $\hat{p}_j^{bs}(\bx), j\in\{1,\ldots, K\} \setminus \{k^*\}$ \orange{and $\hat p_{k^*}^{bs}(\bx)$} from \eqref{eqn:allprob2b}, where $k^*$ is the predefined baseline class. Define the series of weights $0 = \pi_1<\pi_2 <\cdots<\pi_{M}< \pi_{M+1} = 1$, and their maximum interval $\theta_\pi=\max_{\epsilon \in \{1,\ldots, M\}}\{ \Delta \pi_{\epsilon} \mid \Delta \pi_{\epsilon} = \pi_{\epsilon +1}-\pi_{\epsilon}\}$. Then the estimators are consistent, i.e., $\hat{p}_j^{bs}(\bx)\rightarrow p_j(\bx)$ for $j\in\{1, \ldots, K\}$ if $\lambda\rightarrow 0$, $\theta_\pi \rightarrow 0$, and $n\rightarrow \infty$.
\end{theorem}
 
\subsection{Enhanced Pairwise Coupling By Baseline Learning}\label{ssec:3.5}
\blue{Interestingly, the idea of baseline learning can be also used to enhance performance of the original pairwise coupling method in \cite{wang_multiclass_2019} by improving its computation time and estimation accuracy. Recall that, the pairwise coupling requires training $\binom{K}{2}$ binary wSVMs to obtain $\hat{q}_{j|(j, j')}(\bx)$ for all possible $(j,j')$ pairs, which is computationally expensive. In the following, we use the baseline learning solutions $\{\hat{q}_{j|(j, k^*)}(\bx), j \neq k^*\}$ to mathematically derive $\hat{q}_{j|(j, j')}(\bx)$ without actually solving the optimization problems. }

\blue{For any $\bx$, we would like to calculate $\hat{q}_{j|(j, j')}(\bx)$ with $\small\{(j,j') \mid 1\le (j,j') \neq k^* \wedge (j<j') \le K\}$}. Based on the definition of pairwise conditional probability in \cite{wang_multiclass_2019}, we have:
$$\hat{q}_{j|(j, k^*)}(\bx) = \frac{\hat{p}_j(\bx)}{\hat{p}_j(\bx) +\hat{p}_{k^*}(\bx)}, ~~ \hat{q}_{j'|(j', k^*)}(\bx) = \frac{\hat{p}_{j'}(\bx)}{\hat{p}_{j'}(\bx) +\hat{p}_{k^*}(\bx)},~~ \hat{q}_{j|(j, j')}(\bx) = \frac{\hat{p}_{j}(\bx)}{\hat{p}_{j}(\bx) +\hat{p}_{j'}(\bx)}.$$ Therefore, from $\hat{q}_{j|(j, k^*)}(\bx)$ and $\hat{q}_{j'|(j', k^*)}(\bx)$, we have the following relationship 
$$\hat{p}_{k^*}(\bx) = \frac{\hat{p}_j(\bx)}{\hat{q}_{j|(j, k^*)}(\bx)} - \hat{p}_j(\bx) = \frac{\hat{p}_{j'}(\bx)}{\hat{q}_{j'|(j', k^*)}(\bx)} - \hat{p}_{j'}(\bx).$$ 
By applying simple algebraic transformation, we can estimate $q_{j|(j,j')}$ as follows
\begin{equation}\label{probrecons}
    \hat{q}_{j|(j, j')}(\bx) = \frac{\hat{q}_{j|(j, k^*)}(\bx) - \hat{q}_{j|(j, k^*)}(\bx)\hat{q}_{j'|(j', k^*)}(\bx)}{\hat{q}_{j|(j, k^*)}(\bx) + \hat{q}_{j'|(j', k^*)}(\bx) - 2{\hat{q}_{j|(j, k^*)}(\bx)}\hat{q}_{j'|(j', k^*)}(\bx)}.
\end{equation}
 After reconstructing the pairwise conditional probabilities, we can estimate the class probabilities by dynamically selecting the \orange{best} baseline class $k$ for each data point, \orange{by implementing the optimization algorithm of pairwise coupling method discussed in Section \ref{sec:simu}}, which enjoys the linear time complexity. This new type of pairwise coupling enhanced by baseline learning, using $k^*$ chosen \orange{either} by Method 1 or Method 2, is denoted as BP1-SVM and BP2-SVM, respectively. In  Sections \ref{sec:simu} and \ref{sec:real-w}, we will evaluate their performance and make comparisons with \orange{the enhanced implementation of the pairwise coupling} and the baseline learning.

\section{New Learning Scheme: One-vs-All Learning (OVA-SVM)}
\label{sec:4}

\subsection{Methodology}\label{ssec:4.1}
Both of \orange{our enhanced implementation of the pairwise coupling} in \cite*{wang_multiclass_2019} and the baseline learning require the choice of a baseline class $k$, either adaptively for each data point or using a common baseline class $k^*$. In this section, we develop another scheme which does not require choosing the baseline class at all. The new method employs the One-vs-All (OVA) idea, a simple yet effective divide-and-conquer technique for decomposing a large multiclass problem into multiple \orange{smaller and simpler} binary problems. The OVA has shown competitive performance in real-world applications \citep{Rifkin04indefense}. 

The new procedure is proposed as follows. First, we decompose the multiclass problem into $K$ binary problems by training trains class $j$ vs class``not $j$'', where class ``not $j$'' combines all the data points not belonging to class $j$. Second, we fit a weighted SVMs binary classifier separately for the above binary problem for each $j=1, \ldots, K$. Third, we aggregate all the decision rules to construct the class probability estimators for $p_j$'s. We refer this method as OVA-SVM, which is simple to implement and has robust performance. Since the OVA-SVM needs to train unbalanced binary problems, involved with the full data set, its computational cost is higher than the baseline learning. We will perform complexity analysis in Section \ref{sec:algcomplx}.

\subsection{Computation Algorithm}\label{ssec:4.2}
The following is the algorithm to implement the OVA-SVM:
\medskip
\begin{enumerate}[labelwidth=1.5cm,labelindent=10pt,leftmargin=1.7cm,label=\bfseries Step \arabic*.,align=left]
\item[Step 1:] \blue{For each $j$, create class ``not $j$'' by combining all the data points not in class $j$; label the class as $A_j$.} 

\item[Step 2:] 
Define a function $R_{j, A_j}(y)=1$ if $y=j$; and $R_{j, A_j}(y)=-1$ if $y=A_j$. \orange{Fit the kernel binary wSVMs for the class pair $(j, A_j)$ by solving the optimization problem}
         \begin{equation}\label{obj7OVA-SVM}
                \min_{f\in \mathcal{H}_\mathbf{K}}\enskip \frac{1}{n}\Big[(1-\pi_m)\sum_{y_i=j}L(R_{j,A_j}(y_i)f(\bx_i))+\pi_m\sum_{y_i=A_j}L(R_{j, A_j}(y_i)f(\bx_i))\Big]+\lambda J(f)
            \end{equation}
 over a a series \orange{$0 <\pi_2<\cdots<\pi_{M}<1$. Denote the solution to \eqref{obj7OVA-SVM} as $\hat f_{j, A_j, \pi_m}(\cdot)$ for each $j$ and $m$. Furthermore, assume $\hat f_{j, k^*, \pi_1 = 0}(\cdot) >0$ and $\hat f_{j, k^*, \pi_{M+1}=1}(\cdot) < 0$.}

\item[Step 3:] For each class $j$, compute its estimated class probability as
 \begin{equation}\label{eqn:argprobOVA-SVM1}
    \hat{p}_j(\bx)= \frac{1}{2}\left[\arg\min_{\pi_m}\{\hat f_{j, A_j, \pi_m}(\bx) < 0\}+\arg\max_{\pi_m}\{\hat f_{j, A_j, \pi_m}(\bx) > 0\}\right],  \quad  \forall\bx\in\mathcal{S}.
\end{equation}
\item[Step 4:] \blue{Repeat Steps 1-3 and obtain $\hat{p}_j(\bx)$ for $j = 1, \ldots, K$}. 
\end{enumerate}

\medskip
\noindent
To predict the class label for any given input $\bx$, we \orange{apply} $\hat{y} = \arg\max_{1 \le j \le K}\hat{p}_{j}(\bx)$. \blue{At 
Step 2,} $\lambda$ needs to selected adaptively for optimal performance. \orange{We} will discuss the parameter tuning strategy in Section \ref{sec:tuning}.

\subsection{Theory}\label{ssec:4.3}

\cite*{WSL2008} show that the binary wSVMs probability estimator is asymptotically consistent, i.e., $\hat p_{+1}(\bx)$ asymptotically converges to $p_{+1}(\bx)$ as $n \rightarrow \infty$, when the functional space $\mathcal{F}$ is sufficiently rich and and $\lambda$ is properly chosen. For the OVA-SVM approach, the class probability estimator $\hat{p}_j(\bx)$ for each class $j\in \{1,\ldots, K\}$ is solely determined by training the classifier $\hat f_{j, A_j, \pi_m}(\cdot)$ by solving \eqref{obj7OVA-SVM}. Theorem \ref{thm2} below shows the OVA-SVM probability estimators given by \eqref{eqn:argprobOVA-SVM1} are asymptotically consistent for the true class probabilities $p_j(\bx)$'s. The proof is similar to that of Theorems 2 and 3 in \cite*{WSL2008} and thus omitted.

\begin{theorem}\label{thm2}
Consider the $K$-class classification problem using the kernel weighted SVM. Denote the class probabilities estimator
from \eqref{eqn:argprobOVA-SVM1} as $\hat{p}_j(\bx)$ for $j\in\{1,\ldots, K\}$. Define the series of weights $\{0 = \pi_1<\pi_2 <\cdots<\pi_{M}< \pi_{M+1} = 1\}$, and their maximum interval $\theta_\pi=\max_{\epsilon \in \{1,\ldots, M\}}\{ \Delta \pi_{\epsilon} \mid \Delta \pi_{\epsilon} = \pi_{\epsilon +1}-\pi_{\epsilon}\}$. Then the estimators are consistent, i.e., $\hat{p}_j(\bx)\rightarrow p_j(\bx)$ for $j=\{1, \ldots, K\}$ if $\lambda\rightarrow 0$, $\theta_\pi \rightarrow 0$, and the sample size $n\rightarrow \infty$.
\end{theorem}

\section{Complexity Analysis}\label{sec:algcomplx}
We perform rigorous analysis on the computational complexity of a variety of divide-and-conquer learning schemes for multiclass wSVMs, including the pairwise coupling \citep{wang_multiclass_2019}, the proposed baseline learning in Section \ref{sec:3.1}, and the proposed OVA-SVM in Section \ref{sec:4}.

Without loss of generality, assume the $K$ classes are balanced, i.e., $n_{j} = n$ for any $1\le j \le K$. When we randomly split the data as the training and tuning sets, we also keep the class size balanced. For each class pair $\{(j,j') \mid 1 \le j \neq j' \le K\}$, denote the training set and the tuning set as $\mathbb{S}_{train}$ and $\mathbb{S}_{tune}$, and their corresponding sample size as $n_{train}$ and $n_{tune}$, respectively. 

For each binary wSVMs classifier, the optimization problem is solved by Quadratic Programming (QP) using the training set $\mathbb{S}_{train}$. In R, we employ the \texttt{quadprog} package, which uses the active set method and has the polynomial time complexity $\sim O(n_{train}^3)$ \citep{ye_extension_1989,cairanop2013}. The time complexity for label prediction on the tuning set of size $n_{tune}$ by the weighted SVMs classifier is $O(n_{tune}n_{train})$. To train $O(m)$ sequence of wSVMs for pairwise class probability estimation with $n_{tune}$ data points, the time complexity is: $O(m)[O(n_{train}^3) + O(n_{tune}n_{train})] + n_{tune} O(m\log{m})$. During the parameter tuning step, let  $n_{par}$ be the number of tuning parameters and $\{\gamma_p \mid p \in \{1,\ldots,n_{par}\}\} $ be the number of grid points. Then the time complexity for computing the pairwise conditional probability estimators is:
\begin{equation}\label{paircomplx}
    (\prod_{p=1}^{n_{par}}\gamma_p)\left[O(m)(O(n_{train}^3) + O(n_{train}n_{tune})) + n_{tune} O(m\log{m})\right].
\end{equation}

In the following, we focus on the RBF kernel with two hyperparameters $\{\lambda,\sigma\}$, and the number of grid points are $\gamma_\lambda = \gamma_1$ and $\gamma_\sigma = \gamma_2$. We have $m$ as constant and $n_{train} = n_{tune} = n$. Then \eqref{paircomplx} is reduced as  $O(\gamma_1\gamma_2n^{3})$. For the pairwise coupling method, we need to compute $O(K^2)$ pairwise conditional probability estimators, so the total time complexity is $O(\gamma_1\gamma_2K^2n^{3})$. 

For the baseline learning, we first choose the baseline class $k^*$, then compute $O(K)$ pairwise conditional class probabilities. The B1-SVM method chooses $k^*$ as the largest class, and its time complexity is $O(Kn) + O(\gamma_1\gamma_2Kn^{3}) = O(\gamma_1\gamma_2Kn^{3})$. The B2-SVM method selects $k^*$ by measuring the median class distance. Since the time complexity of computing $K$-class $\mathcal{D}_{cp}$ and $\mathcal{D}_{bc}$ is $O(Kn^2)$, we need $O(Kn^2)$ to obtain $k^*$. Then the time complexity for B2-SVM is $O(Kn^2) + O(\gamma_1\gamma_2Kn^{3}) = O(\gamma_1\gamma_2Kn^{3})$, which is the same as B1-SVM.

Now we consider the time complexity of the One-vs-All method (OVA-SVM), where we calculate $O(K)$ pairwise conditional class probabilities. Assume that $n_{train} = n_{tune} = 0.5Kn$. From \eqref{paircomplx}, we have $ O(\gamma_1\gamma_2K^{3}n^{3})$, and so the total time complexity for OVA-SVM is $O(\gamma_1\gamma_2K^{4}n^{3})$.

In summary, for $K$-class problems with the wSVMs, for fixed $\gamma_1$, $\gamma_2$, $m$, and $n$, the time complexity for the baseline learning methods (B1-SVM and B2-SVM) is $\propto O(K)$, for the pairwise coupling is $\propto O(K^2)$, and for OVA-SVM is $\propto O(K^{4})$. The proposed baseline method is linear in $K$, which has the lowest computational cost among all of the divide-and-conquer wSVM \orange{methods}. The above analysis is based on a balanced scenario. For the unbalanced case, the analysis is more complex mathematically, but the conclusion remains the same.

\pagebreak
\section{Parameter Tuning}\label{sec:tuning}
The proposed kernel learning method involves two tuning parameters, 
$\lambda>0$ and $\sigma \in \realset$ (if an RBF kernel is used). Their values are critical to assure the optimal finite sample performance of the wSVM. To choose proper values for these tuning parameters, we randomly split the data into two equal parts: the training set $\mathbb{S}_{train}$ used to train the wSVM at \eqref{obj8ba} \orange{or \eqref{obj7OVA-SVM}} for a given weight $\pi$ and fixed $(\lambda, \sigma)$, and the tuning set used to select the \orange{optimal values of $\{\lambda, \sigma\}$}. 

For a fixed $\{\lambda, \sigma\}$, we train $M-1$ classifiers $\hat{f}_{\pi_{\epsilon}}^{\lambda,\sigma}(\bx)$ for $\pi_\epsilon \in \{\frac{j-1}{M} \mid j=2, \ldots, M\}$ $\mathbb{S}_{train}$ to obtain the pairwise conditional probability \orange{estimators} $\hat{q}_{j|(j, j')}^{\lambda,\sigma}$, where $j'$ is the common baseline class $k^*$ in the baseline learning and $A_j$ for the OVA-SVM. We propose choosing the tuning parameters using a grid search, based on the prediction performance of $\hat{q}^{\lambda,\sigma}$'s on the tuning set $\mathbb{S}_{tune}$. Assume the true pairwise conditional probabilities are $q_{j|(j, j')}(\bx)$. To quantify the estimation accuracy of $\hat{q}$'s, we use the generalized Kullback–Leibler (GKL) loss function
\begin{eqnarray}    
&&\mbox{GKL}(q_{j|(j, j')}, \hat {q}^{\lambda,\sigma}_{j|j, j')})\label{eq:gkl1}\\
&=&\mathlarger{\mathop{\mathbb{E}}}_{\bX\sim \mathcal{P}}\left[q_{j|(j, j')}(\bX)\log \frac{q_{j|(j, j')}(\bX)}{\hat q^{\lambda,\sigma}_{j|(j, j')}(\bX)}+(1-q_{j|(j, j')}(\bX))\log \frac{1-q_{j|(j, j')}(\bX)}{1-\hat q^{\lambda,\sigma}_{j|(j,j')}(\bX)}\right]\nonumber    \\
&=&\mathlarger{\mathop{\mathbb{E}}}_{\bX\sim \mathcal{P}}\left[q_{j|(j, j')}(\bX)\log {q_{j|(j, j')}(\bX)}+(1-q_{j|(j, j')}(\bX))\log( {1-q_{j|(j, j')}(\bX)})\right]\label{eq:gkl2}    \\ 
&\;&- \mathlarger{\mathop{\mathbb{E}}}_{\bX\sim \mathcal{P}}\left[q_{j|(j, j')}(\bX)\log {\hat q^{\lambda,\sigma}_{j|(j, j')}(\bX)}+(1-q_{j|(j, j')}(\bX))\log( {1-\hat q^{\lambda,\sigma}_{j|(j, j')}(\bX)})\right].\label{eq:gkl3}
\end{eqnarray}
The term \eqref{eq:gkl2} is unrelated to $\hat q^{\lambda,\sigma}_{j|(j, j')}(\bX)$, so only the term \eqref{eq:gkl3} is essential. Since the true values of $q_{j|(j, j')}(\bX)$ are unknown in practice, we need to approximate GKL empirically based on the tuning data. Set a function $R_{j, j'}(Y)=1$ for $y_i = j$; $=-1$ for $y_i = j'$, which satisfies $\mathlarger{\mathop{\mathbb{E}}}\left[\frac{1}{2}(R_{j, j'}(Y)+1)|\bX, Y\in \{j, j'\}\right] = q_{j|(j, j')}(\bX)$, so a computable proxy of GKL is given by
\begin{eqnarray}    
&&\mbox{EGKL}(\hat {q}^{\lambda,\sigma}_{j|j, j')})\label{eq:egkl}\\
&=&-\frac{1}{2n_{jj'}}\sum_{\forall i: y_i=j \vee j'}\left[(1+R_{j, j'}(y_i))\log \hat q^{\lambda,\sigma}_{j|(j, j')}(\bx_i)
+ (1-R_{j, j'}(y_i))\log (1-\hat q^{\lambda,\sigma}_{j|(j, j')}(\bx_i))\right]. \nonumber    
\end{eqnarray}
The \orange{optimal values of $\{\lambda, \sigma\}$} are chosen by minimizing the EGKL.

\section{Numerical Studies}\label{sec:simu}
\subsection{Experiment Setup}
We evaluate the performance of the baseline learning (denoted by B-SVM) and the OVA-SVM (denoted by A-SVM) schemes, and the pairwise coupling enhanced by the baseline learning (denoted by BP-SVM). We compare them with \orange{our enhanced implementation of the pairwise coupling algorithm in \cite*{wang_multiclass_2019} to dynamically chooses the best baseline class for each data point to be the class $k$ that has the maximum voting for larger pairwise conditional probabilities} (denoted by P-SVM). All of these methods are implemented in R, and the tuning parameters are chosen using EGKL based on a tuning set. In addition, we include five popular benchmark methods in the literature: multinomial logistic regression (MLG), multiclass linear discriminant analysis (MLDA), classification tree \citep[TREE,][]{cart84}, random forest \citep[RF,][]{ho1995random}, and XGBoost \citep[XGB,][]{Chenxgb2016}, which are implemented in R with hyperparameters tuned by cross-validation. All of the numerical experiments are performed on the High Performance Computing (HPC) cluster at the University of Arizona, with Lenovo's Nextscale M5 Intel Haswell V3 28 core processors and 18GB memory.

In all the simulation examples, we set $\Card{\mathbb{S}_{train}} = \Card{\mathbb{S}_{tune}} = n$.
The weights are specified as $\pi_{j} = (j-1)/M, j = \{1, \ldots, M+1\}$, where $M$ is a preset value to control estimation precision. Following \cite{WSL2008}, we set $M= \floor{\sqrt{n}}$. We use the RBF kernel and select the optimal values of $\{\lambda,\sigma\}$ using GKL or EGKL based on a grid search. The range of $\sigma$ is $\{ \sigma_M/4, 2 \sigma_M/4, 3 \sigma_M/4, 4 \sigma_M/4, 5 \sigma_M/4, 6 \sigma_M/4\}$, where $\sigma_M=\mbox{median}_{y_s\ne y_t}\norms{{\bx_s-\bx_t}}_{2}$ is the median pairwise Euclidean distance between $K$ classes \citep{WuZhaLiu2010}, and $\lambda \in \{5.5 \times 10^{-8}, 1 \times 10^{-8}, 5.5 \times 10^{-7},  \ldots, 5.5 \times 10^{+7}, 1 \times 10^{+8}\}$. We report the results of both GKL and EGKL, showing that they give a comparable performance. In real applications, since GKL is not available, we will use EGKL for tuning. 

To evaluate the estimation accuracy of the probability estimator $\hat{p}_j$'s, we further  generate a test set $\mathbb{S}_{test} = \{(\tilde \bx_i, \tilde y_i), i=1, \ldots, \tilde n\}$, of the size $\Card{\mathbb{S}_{test}} = \tilde{n}$. The following metrics are used to quantify the performance of different procedures: (i) 1-norm error $\frac{1}{\tilde n}\sum_{i=1}^{\tilde n}\sum_{j=1}^{K} |\hat{p}_j(\tilde\bx_i)-p_j(\tilde\bx_i)|$; (ii) 2-norm error $\frac{1}{\tilde n}\sum_{i=1}^{\tilde n}\sum_{j=1}^{K}(\hat{p}_j(\tilde\bx_i)-p_j(\tilde\bx_i))^2$; (iii) the EGKL loss $\frac{1}{\tilde n}\sum_{i=1}^{\tilde n}\sum_{j=1}^{K} p_j(\tilde \bx_i)\log\frac{p_j(\tilde \bx_i)}{\hat{p}_j(\tilde\bx_i)}$; and (iv) the GKL loss $\frac{1}{\tilde n}\sum_{i=1}^{\tilde n}\sum_{j=1}^{K}\left[ p_j(\tilde \bx_i)\log\frac{p_j(\tilde \bx_i)}{\hat{p}_j(\tilde\bx_i)} + (1-p_j(\tilde \bx_i))\log\frac{(1-p_j(\tilde \bx_i))}{(1-\hat{p}_j(\tilde\bx_i))}\right]$.
To evaluate the classification accuracy of the learning methods, we report two misclassification rates based on the maximum probability and max voting algorithms. 

We consider four examples, including two linear cases (Examples 1-2) and two nonlinear cases (Examples 3-4). The number of classes $K\in\{ 5, 7, 9\}$. MLG and MLDA are oracles methods for Examples 1-2, but not for Examples 3-4. Figure 1 illustrates the scatter plots of the training data points for Examples 1-4, along with the optimal (Bayes) decision boundary shown in solid lines. For all the examples, $n=500$ and $\tilde{n} = 10,000$. For each example, we run $N_{r}=100$ Monte Carlo simulations and report the average performance measurement with standard error.

\subsection{Linear Examples}

\textbf{Example 1 ($K=7$, linear)}. The data are generated as follows: (i) Generate $Y$ uniformly from $\{1, 2, 3, 4,5,6,7\}$; (ii) Given $Y=y$, generate $\bX$ from the bivariate normal distribution $N({\bmu}(y),\mathbf{\Sigma})$, where $\bmu(y)=(1.5\cos(2y\pi/7),1.5\sin(2y\pi/7))^{\small\mathrm{T}}$, $\mathbf{\Sigma}=1.2^2\mathbf{I}_2$.

\medskip
\noindent
\textbf{Example 2 ($K=9$, linear)}. The data are generated as follows: (i) Generate $Y$ uniformly from $\{1, 2, 3, 4, 5,6,7,8,9\}$; (ii) Given $Y=y$, generate $\bX$ from the bivariate normal distribution $N({\bmu}(y),\mathbf{\Sigma})$, where $\bmu(y)=(2.5\cos(2y\pi/9),2.5\sin(2y\pi/9))^{\small\mathrm{T}}$, $\mathbf{\Sigma}=1.5^2\mathbf{I}_2$.

\medskip
\noindent
Examples 1-2 are two linear examples used to show that the baseline methods can handle a large class size $K \ge 5$ with competitive performance and the best running time. Table 1 summarizes the performance of the baseline learning methods (B1-SVM, B2-SVM), the pairwise coupling enhanced by the baseline learning (BP1-SVM, BP2-SVM), respectively, based on two methods of choosing the baseline class $k^*$, the \orange{OVA-SVM method (A-SVM)}, the pairwise coupling \orange{ with dynamic baseline class choosing for each data point} (P-SVM), and five benchmark methods: MLG (Oracle-1), MLDA (Oracle-2), TREE, RF and XGBoost. We consider 2 different class sizes and report seven performance measures: running time, 1-norm error, 2-norm error, EGKL loss, GKL loss, and the misclassification rate for classification. The standard error (SE), \orange{as calculated by $\sigma_{\bar{v}} = \sigma_{\bv}/\sqrt{N_{r}}$}, is in the parenthesis.

\begin{figure}[H]
\centering
\caption{The scatter plots of the training data for Examples 1-4. Different symbols are for different classes, and the optimal decision boundaries (Bayes rules) are solid lines.}
     \includegraphics[max size={1\textwidth}{\textheight}]{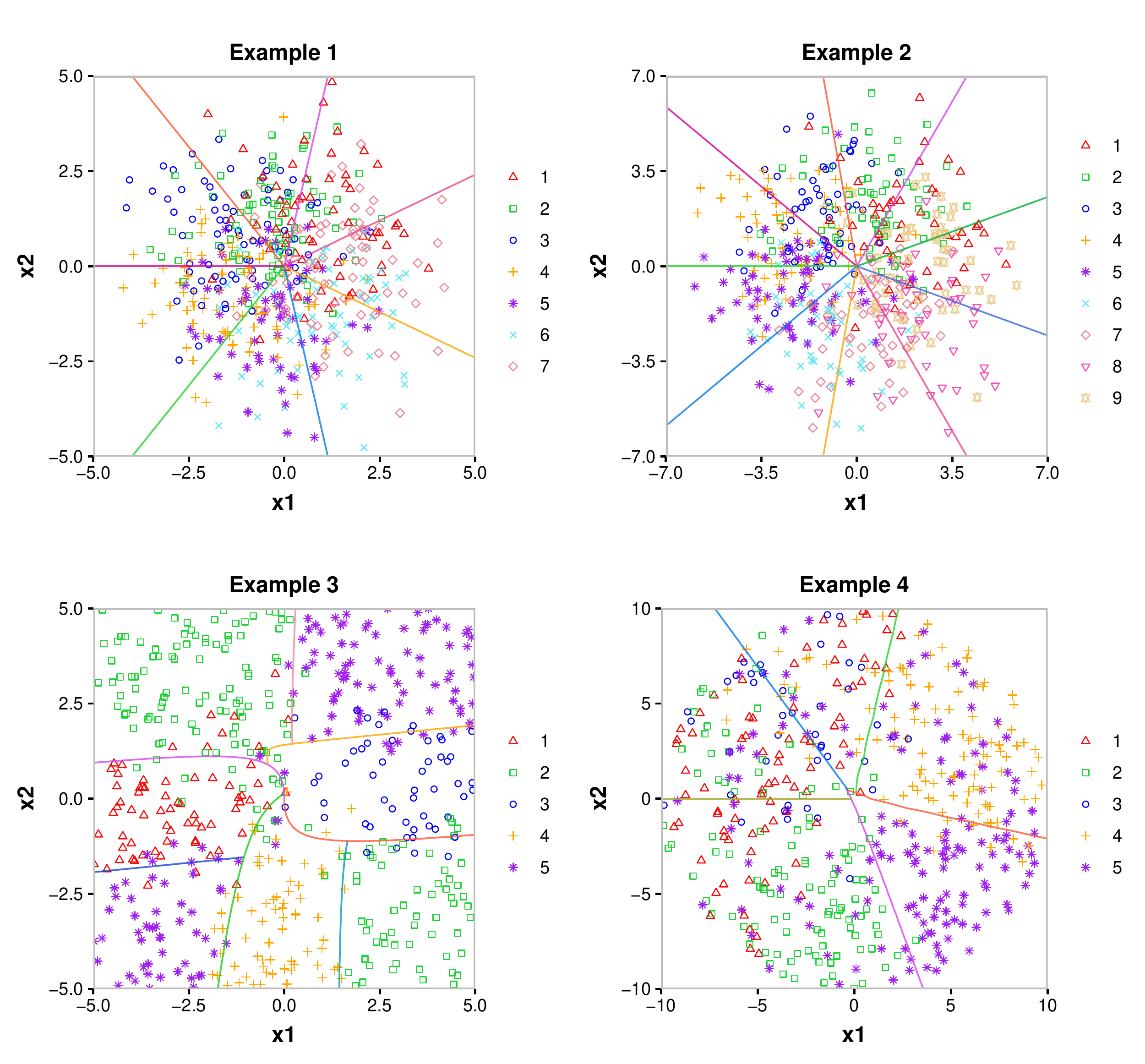}
              \phantomsubcaption
        \label{fig:1}
\end{figure}

\begin{table}[H]
\caption{Performance measure on the test set for Example 1 and Example 2.}
\centering
\scalebox{0.635}{
\begin{tabular}{c|cc|cc|c|c|ccccc} 
\hline
\multicolumn{12}{c}{\textbf{Example 1 (Seven-class Linear Example, MLG and MLDA as Oracle, denoted as Oracle-1 and Oracle-2)}}                 \\ 
\hline
\textbf{EGKL} & \textbf{B1-SVM} & \textbf{BP1-SVM} & \textbf{B2-SVM} & \textbf{BP2-SVM} & \textbf{A-SVM} & \textbf{P-SVM} & \textbf{Oracle-1} & \textbf{Oracle-2} & \textbf{TREE} & \textbf{RF} & \textbf{XGB}  \\ 
\hline
Time          & 0.9 (0.0)       & 0.9 (0.0)        & 0.8 (0.0)       & 0.8 (0.0)        & 30.7 (0.1)     & 2.8 (0.0)      & 0.0 (0.0)    & 0.0 (0.0)     & 0.4 (0.0)     & 1.8 (0.0)   & 1.8 (1.1)     \\ 
$L_1$            & 47.9 (0.2)      & 47.9 (0.2)       & 48.0 (0.3)      & 48.0 (0.3)       & 41.8 (0.2)     & 46.5 (0.1)     & 36.5 (0.2)   & 36.0 (0.1)    & 56.3 (0.2)    & 72.0 (0.1)  & 95.2 (0.2)    \\
$L_2$            & 9.5 (0.1)       & 9.5 (0.1)        & 9.8 (0.1)       & 9.8 (0.1)        & 8.5 (0.1)      & 9.1 (0.0)      & 8.2 (0.1)    & 8.1 (0.0)     & 11.7 (0.1)    & 19.4 (0.1)  & 35.6 (0.1)    \\
EGKL          & 31.7 (1.9)      & 31.7 (1.9)       & 33.3 (1.9)      & 33.3 (1.9)       & 26.7 (1.3)     & 30.2 (1.2)     & 22.5 (1.2)   & 22.1 (1.2)    & Inf (NaN)     & Inf (NaN)   & 124.6 (2.6)   \\
GKL           & 25.1 (0.7)      & 25.1 (0.7)       & 26.9 (1.2)      & 26.9 (1.2)       & 15.1 (0.4)     & 24.6 (0.4)     & 14.8 (0.4)   & 14.4 (0.3)    & Inf (NaN)     & Inf (NaN)   & 162.0 (1.4)   \\ 
TE1           & 60.7 (0.2)      & 60.5 (0.2)       & 60.9 (0.2)      & 60.7 (0.2)       & 60.0 (0.1)     & 60.3 (0.1)     & 57.6 (0.1)   & 57.7 (0.1)    & 61.9 (0.1)    & 64.7 (0.1)  & 65.7 (0.1)    \\
TE2           & NA (NA)         & 60.3 (0.2)       & NA (NA)         & 60.5 (0.2)       & NA (NA)        & 60.3 (0.1)     & NA (NA)      & NA (NA)       & NA (NA)       & NA (NA)     & NA (NA)       \\ 
$k^*$          & 1               & 1                & 5               & 5                & NA (NA)        & NA (NA)        & NA (NA)      & NA (NA)       & NA (NA)       & NA (NA)     & NA (NA)       \\ 
\hline
\textbf{GKL}  & \textbf{B1-SVM} & \textbf{BP1-SVM} & \textbf{B2-SVM} & \textbf{BP2-SVM} & \textbf{A-SVM} & \textbf{P-SVM} & \textbf{Oracle-1} & \textbf{Oracle-2} & \textbf{TREE} & \textbf{RF} & \textbf{XGB}  \\ 
\hline
Time          & 0.9 (0.0)       & 0.9 (0.0)        & 0.8 (0.0)       & 0.8 (0.0)        & 30.8 (0.1)     & 2.8 (0.0)      & 0.0 (0.0)    & 0.0 (0.0)     & 0.4 (0.0)     & 1.8 (0.0)   & 1.8 (1.1)     \\ 
$L_1$            & 48.4 (0.3)      & 48.4 (0.3)       & 48.5 (0.3)      & 48.5 (0.3)       & 42.2 (0.2)     & 46.9 (0.2)     & 36.5 (0.2)   & 36.0 (0.1)    & 56.3 (0.2)    & 72.0 (0.1)  & 95.2 (0.2)    \\
$L_2$            & 10.4 (0.1)      & 10.4 (0.1)       & 10.3 (0.1)      & 10.3 (0.1)       & 9.2 (0.0)      & 9.8 (0.1)      & 8.2 (0.1)    & 8.1 (0.0)     & 11.7 (0.1)    & 19.4 (0.1)  & 35.6 (0.1)    \\
EGKL          & 31.8 (2.4)      & 31.8 (2.4)       & 33.9 (2.6)      & 33.9 (2.6)       & 26.8 (1.3)     & 30.3 (1.2)     & 22.5 (1.2)   & 22.1 (1.2)    & Inf (NaN)     & Inf (NaN)   & 124.6 (2.6)   \\
GKL           & 25.3 (1.1)      & 25.3 (1.1)       & 26.8 (0.9)      & 26.8 (0.9)       & 15.2 (0.4)     & 24.6 (0.4)     & 14.8 (0.4)   & 14.4 (0.3)    & Inf (NaN)     & Inf (NaN)   & 162.0 (1.4)   \\ 
TE1           & 60.8 (0.2)      & 60.6 (0.2)       & 61.2 (0.2)      & 60.9 (0.2)       & 59.9 (0.1)     & 60.3 (0.2)     & 57.6 (0.1)   & 57.7 (0.1)    & 61.9 (0.1)    & 64.7 (0.1)  & 65.7 (0.1)    \\
TE2           & NA (NA)         & 60.4 (0.2)       & NA (NA)         & 60.8 (0.3)       & NA (NA)        & 60.3 (0.2)     & NA (NA)      & NA (NA)       & NA (NA)       & NA (NA)     & NA (NA)       \\ 
$k^*$          & 1               & 1                & 5               & 5                & NA (NA)        & NA (NA)        & NA (NA)      & NA (NA)       & NA (NA)       & NA (NA)     & NA (NA)       \\ 
\hline
\multicolumn{12}{c}{\textbf{Example 2 (Nine-class Linear Example, MLG and MLDA as Oracle, denoted as Oracle-1 and Oracle-2)}}                                                                                                            \\ 
\hline
\textbf{EGKL} & \textbf{B1-SVM} & \textbf{BP1-SVM} & \textbf{B2-SVM} & \textbf{BP2-SVM} & \textbf{A-SVM} & \textbf{P-SVM} & \textbf{Oracle-1} & \textbf{Oracle-2} & \textbf{TREE} & \textbf{RF} & \textbf{XGB}  \\ 
\hline
Time          & 0.8 (0.0)       & 0.8 (0.0)        & 0.8 (0.0)       & 0.8 (0.0)        & 31.6 (0.2)     & 3.4 (0.0)      & 0.0 (0.0)    & 0.0 (0.0)     & 0.4 (0.0)     & 1.8 (0.0)   & 7.2 (2.8)     \\ 
$L_1$            & 46.1 (0.3)      & 46.1 (0.3)       & 47.0 (0.3)      & 47.0 (0.3)       & 43.5 (0.1)     & 45.2 (0.1)     & 43.5 (0.2)   & 42.7 (0.2)    & 64.0 (0.2)    & 75.2 (0.1)  & 99.8 (0.1)    \\
$L_2$            & 12.4 (0.1)      & 12.4 (0.1)       & 12.3 (0.1)      & 12.3 (0.1)       & 11.1 (0.1)     & 12.0 (0.0)     & 10.6 (0.1)   & 10.4 (0.1)    & 14.5 (0.1)    & 21.7 (0.1)  & 40.0 (0.1)    \\
EGKL          & 32.0 (2.4)      & 32.0 (2.4)       & 32.2 (2.2)      & 32.2 (2.2)       & 27.0 (1.6)     & 31.6 (1.2)     & 26.9 (1.4)   & 25.9 (1.3)    & Inf (NaN)     & Inf (NaN)   & 145.2 (2.8)   \\
GKL           & 22.6 (1.1)      & 22.6 (1.1)       & 22.5 (1.2)      & 22.5 (1.2)       & 18.5 (0.6)     & 21.7 (0.5)     & 18.1 (0.4)   & 16.9 (0.4)    & Inf (NaN)     & Inf (NaN)   & 187.1 (1.8)   \\ 
TE1           & 61.2 (0.2)      & 60.4 (0.2)       & 61.6 (0.3)      & 60.9 (0.3)       & 60.1 (0.1)     & 60.9 (0.1)     & 57.4 (0.1)   & 57.3 (0.1)    & 62.5 (0.1)    & 64.2 (0.1)  & 65.3 (0.1)    \\
TE2           & NA (NA)         & 60.3 (0.2)       & NA (NA)         & 60.6 (0.3)       & NA (NA)        & 60.9 (0.1)     & NA (NA)      & NA (NA)       & NA (NA)       & NA (NA)     & NA (NA)       \\ 
$k^*$          & 6               & 6                & 8               & 8                & NA (NA)        & NA (NA)        & NA (NA)      & NA (NA)       & NA (NA)       & NA (NA)     & NA (NA)       \\ 
\hline
\textbf{GKL}  & \textbf{B1-SVM} & \textbf{BP1-SVM} & \textbf{B2-SVM} & \textbf{BP2-SVM} & \textbf{A-SVM} & \textbf{P-SVM} & \textbf{Oracle-1} & \textbf{Oracle-2} & \textbf{TREE} & \textbf{RF} & \textbf{XGB}  \\ 
\hline
Time          & 0.8 (0.0)       & 0.8 (0.0)        & 0.8 (0.0)       & 0.8 (0.0)        & 31.7 (0.2)     & 3.4 (0.0)      & 0.0 (0.0)    & 0.0 (0.0)     & 0.4 (0.0)     & 1.8 (0.0)   & 7.2 (2.8)     \\ 
$L_1$            & 46.1 (0.5)      & 46.1 (0.5)       & 47.1 (0.4)      & 47.1 (0.4)       & 43.1 (0.1)     & 45.2 (0.3)     & 43.5 (0.2)   & 42.7 (0.2)    & 64.0 (0.2)    & 75.2 (0.1)  & 99.8 (0.1)    \\
$L_2$            & 12.6 (0.2)      & 12.6 (0.2)       & 12.8 (0.1)      & 12.8 (0.1)       & 11.6 (0.1)     & 12.5 (0.1)     & 10.6 (0.1)   & 10.4 (0.1)    & 14.5 (0.1)    & 21.7 (0.1)  & 40.0 (0.1)    \\
EGKL          & 32.0 (3.5)      & 32.0 (3.5)       & 32.9 (3.1)      & 32.9 (3.1)       & 27.1 (1.6)     & 31.6 (1.4)     & 26.9 (1.4)   & 25.9 (1.3)    & Inf (NaN)     & Inf (NaN)   & 145.2 (2.8)   \\
GKL           & 22.5 (3.2)      & 22.5 (3.2)       & 22.3 (1.9)      & 22.3 (1.9)       & 18.1 (0.6)     & 21.5 (1.0)     & 18.1 (0.4)   & 16.9 (0.4)    & Inf (NaN)     & Inf (NaN)   & 187.1 (1.8)   \\ 
TE1           & 62.0 (0.3)      & 61.9 (0.3)       & 62.0 (0.3)      & 61.8 (0.3)       & 60.7 (0.1)     & 61.3 (0.3)     & 57.4 (0.1)   & 57.3 (0.1)    & 62.5 (0.1)    & 64.2 (0.1)  & 65.3 (0.1)    \\
TE2           & NA (NA)         & 61.3 (0.3)       & NA (NA)         & 61.3 (0.3)       & NA (NA)        & 61.3 (0.2)     & NA (NA)      & NA (NA)       & NA (NA)       & NA (NA)     & NA (NA)       \\ 
$k^*$          & 6               & 6                & 8               & 8                & NA (NA)        & NA (NA)        & NA (NA)      & NA (NA)       & NA (NA)       & NA (NA)     & NA (NA)       \\
\hline
\end{tabular}}
\end{table}

\hfill \break
\hfill \break
\textbf{TABLE NOTE:} The running time is measured in minutes. $L_1$ (the 1-norm error), $L_2$ (the 2-norm error), EGKL (the EGKL loss), GKL (the GKL loss), TE1 (the misclassification rate based on max probability), TE2 (the misclassification rate based on max voting), multiplied by 100 for both mean and standard derivation in parenthesis. Baseline class $k^*$ in baseline methods are calculated by statistical mode.  MLG: Multinomial logistic regression; MLDA: Multiclass linear discriminant analysis; TREE: Classification tree; RF: Random forest; XGB: XGBoost. The same explanation applies to the results of all the other examples.

\medskip
We make the following observations. In general, SVM-based methods outperform other methods and are closest to the oracle methods in all cases. Two B-SVM estimators have similar probability estimation performance as the pairwise coupling methods, and they outperform all three benchmark methods besides the oracle procedures (MLG and MLDA), showing a significant gain in computation efficiency when $K$ is large. The BP-SVM methods have no additional running cost since they don't involve additional training, showing better classification accuracy in some cases. Based on data distribution, two B-SVM methods may select different \orange{baseline} class $k^*$, which may give slightly different yet similar results. The OVA-SVM consistently gives the best probability estimation and classification in all methods, however with the most expensive \orange{computational} cost, which agrees with the theoretic complexity analysis. The tuning criteria GKL and EGKL give similar performance, showing that EGKL is a good approximation as GKL for parameter tuning. Noted, the max voting algorithm works only with the full pairwise probability table, so it is not applicable for OVA-SVM and B-SVM, as denoted as NA. The tree based methods, such as classification trees (TREE) and random forest (RF), may incur 0 in class probability estimation for some $\bx$ and classes, resulting "Inf" in Tables 1-4 for both EGKL and GKL, thus the SEs are unavailable and output NaN ("Not A Number") in R. 


\subsection{Nonlinear Examples}
Here we show that the proposed methods can handle complex non-linear problems and offer competitive performance in both probability estimation and classification accuracy. In these examples, the linear methods MLG and MLDA are not the oracle anymore. 

\smallskip

\noindent
\textbf{Example 3 ($K=5$, nonlinear)}. For any $\bx=(x_1, x_2)^{\small\mathrm{T}}$, define
  \begin{eqnarray*}
f_1(\bx)&=&-1.5x_1+0.2x_1^2-0.1x_2^2+0.2\\
f_2(\bx)&=&0.3x_1^2+0.2x_2^2-x_1x_2+0.2\\
f_3(\bx)&=&1.5x_1+0.2x_1^2-0.1x_2^2+0.2\\
f_4(\bx)&=&-0.1x_1^2+0.2x_2^2-1.5x_2+x_1+0.1x_1x_2\\
f_5(\bx)&=&0.1x_1^2+0.1x_2^2 + x_1x_2 - 0.2
\end{eqnarray*}
and $p_j(\bx)= e^{f_j(\bx)}/(\sum_{l=1}^5e^ {f_l(\bx)})$ for $j=1, 2, 3, 4,5$. The data are generated as follows: (i) Generate independent $X_1$ and $X_2$ uniformly from $[-5, 5]$ and $[-5,5]$, respectively; (ii) Given $\bX=\bx$, the label $Y$ takes the value $j$ with the probability $p_j(\bx)$ for $j=1,2,3,4,5$.

\medskip
\noindent
\textbf{Example 4 ($K=5$, nonlinear).} Generate $\bX=(X_1, X_2)^{\small\mathrm{T}}$ uniformly from a disc $\{\bx: x_1^2+x_2^2\leq 100\}$. Define $h_1(\bx)=-3x_1\sqrt{5}+3x_2$, $h_2(\bx)=-3x_1\sqrt{5}-3x_2$,
$h_3(\bx)=x_2\sqrt{3}-1.2x_1$,
$h_4(\bx)=2x_2\sqrt{3}+1.2x_1$, and
$h_5(\bx)=\sqrt{|x_1x_2|+1}$.
Then consider the non-linear transformation $f_j(\bx)= \Phi^{-1}(T_2(h_j(\bx)))$ for $j=1,2,3,4,5$, where $\Phi(\cdot)$ is the cumulative distribution function(cdf) of the standard normal distribution and $T_2(\cdot)$ is the cdf of $t_2$ distribution. Then we set the probability functions as $p_j(\bx)= \exp{f_j(\bx)}/(\sum_{l=1}^5\exp(f_l(\bx)))$ for $j=1,2,3,4,5$. 

\medskip
\medskip
Table 2 summarizes the performance of the proposed methods and their comparison with the pairwise coupling and five benchmark methods. Similar to the linear cases, the proposed OVA-SVM has the best performance among all the methods. The B-SVMs have close performance as pairwise coupling and outperform all the benchmark methods.

\begin{table}[H]
\caption{Performance measure on the test set for Example 3 and Example 4.}
\centering
\scalebox{0.635}{
\begin{tabular}{c|cc|cc|c|c|ccccc} 
\hline

\multicolumn{12}{c}{\textbf{Example 3 (Five-class Non-linear Example, MLG and MLDA are not Oracle)}}                                           \\ 
\hline
\textbf{EGKL} & \textbf{B1-SVM} & \textbf{BP1-SVM} & \textbf{B2-SVM} & \textbf{BP2-SVM} & \textbf{A-SVM} & \textbf{P-SVM} & \textbf{MLG} & \textbf{MLDA} & \textbf{TREE} & \textbf{RF} & \textbf{XGB}  \\ 
\hline
Time          & 1.5 (0.0)       & 1.5 (0.0)        & 0.7 (0.0)       & 0.7 (0.0)        & 11.7 (0.0)     & 2.4 (0.0)      & 0.0 (0.0)    & 0.0 (0.0)     & 0.3 (0.0)     & 1.1 (0.0)   & 0.4 (0.0)     \\ 
$L_1$            & 19.2 (0.3)      & 19.2 (0.3)       & 21.1 (1.2)      & 21.1 (1.2)       & 15.2 (0.1)     & 18.9 (0.1)     & 115.9 (0.1)  & 116.3 (0.1)   & 27.9 (0.2)    & 22.4 (0.1)  & 27.9 (0.1)    \\
$L_2$            & 4.1 (0.2)       & 4.1 (0.2)        & 4.3 (0.7)       & 4.3 (0.7)        & 3.3 (0.1)      & 4.0 (0.1)      & 50.7 (0.1)   & 50.4 (0.1)    & 7.0 (0.1)     & 4.4 (0.1)   & 9.4 (0.1)     \\
EGKL          & 12.5 (0.9)      & 12.5 (0.9)       & 13.7 (1.9)      & 13.7 (1.9)       & 9.2 (0.6)      & 12.2 (0.5)     & 60.7 (0.8)   & 60.8 (0.8)    & Inf (NaN)     & Inf (NaN)   & 28.6 (1.2)    \\
GKL           & 22.6 (0.6)      & 22.6 (0.6)       & 26.9 (2.3)      & 26.9 (2.3)       & 15.6 (0.3)     & 20.8 (0.3)     & 88.5 (0.2)   & 89.7 (0.2)    & Inf (NaN)     & Inf (NaN)   & 49.7 (0.7)    \\ 
TE1           & 16.6 (0.3)      & 16.6 (0.3)       & 16.7 (0.8)      & 16.7 (0.8)       & 16.4 (0.1)     & 16.5 (0.1)     & 62.1 (0.4)   & 61.8 (0.4)    & 17.5 (0.1)    & 17.2 (0.1)  & 18.2 (0.1)    \\
TE2           & NA (NA)         & 16.5 (0.3)       & NA (NA)         & 16.6 (0.8)       & NA (NA)        & 16.6 (0.1)     & NA (NA)      & NA (NA)       & NA (NA)       & NA (NA)     & NA (NA)       \\ 
$k^*$          & 2               & 2                & 4               & 4                & NA (NA)        & NA (NA)        & NA (NA)      & NA (NA)       & NA (NA)       & NA (NA)     & NA (NA)       \\ 
\hline
\textbf{GKL}  & \textbf{B1-SVM} & \textbf{BP1-SVM} & \textbf{B2-SVM} & \textbf{BP2-SVM} & \textbf{A-SVM} & \textbf{P-SVM} & \textbf{MLG} & \textbf{MLDA} & \textbf{Tree} & \textbf{RF} & \textbf{XGB}  \\ 
\hline
Time          & 1.5 (0.0)       & 1.5 (0.0)        & 0.7 (0.0)       & 0.7 (0.0)        & 11.7 (0.0)     & 2.4 (0.0)      & 0.0 (0.0)    & 0.0 (0.0)     & 0.3 (0.0)     & 1.1 (0.0)   & 0.4 (0.0)     \\ 
$L_1$            & 19.0 (0.6)      & 19.0 (0.6)       & 21.4 (1.3)      & 21.4 (1.3)       & 15.9 (0.1)     & 18.7 (0.2)     & 115.9 (0.1)  & 116.3 (0.1)   & 27.9 (0.2)    & 22.4 (0.1)  & 27.9 (0.1)    \\
$L_2$            & 4.2 (0.4)       & 4.2 (0.4)        & 4.4 (0.8)       & 4.4 (0.8)        & 3.2 (0.0)      & 4.1 (0.2)      & 50.7 (0.1)   & 50.4 (0.1)    & 7.0 (0.1)     & 4.4 (0.1)   & 9.4 (0.1)     \\
EGKL          & 12.9 (1.2)      & 12.9 (1.2)       & 13.7 (1.6)      & 13.7 (1.6)       & 8.9 (0.5)      & 11.3 (0.5)     & 60.7 (0.8)   & 60.8 (0.8)    & Inf (NaN)     & Inf (NaN)   & 28.6 (1.2)    \\
GKL           & 22.6 (0.9)      & 22.6 (0.9)       & 26.3 (1.5)      & 26.3 (1.5)       & 15.3 (0.3)     & 20.9 (0.3)     & 88.5 (0.2)   & 89.7 (0.2)    & Inf (NaN)     & Inf (NaN)   & 49.7 (0.7)    \\ 
TE1           & 16.7 (0.4)      & 16.7 (0.4)       & 16.7 (1.0)      & 16.7 (1.0)       & 16.4 (0.1)     & 16.6 (0.1)     & 62.1 (0.4)   & 61.8 (0.4)    & 17.5 (0.1)    & 17.2 (0.1)  & 18.2 (0.1)    \\
TE2           & NA (NA)         & 16.6 (0.4)       & NA (NA)         & 16.6 (1.0)       & NA (NA)        & 16.5 (0.1)     & NA (NA)      & NA (NA)       & NA (NA)       & NA (NA)     & NA (NA)       \\ 
$k^*$          & 2               & 2                & 4               & 4                & NA (NA)        & NA (NA)        & NA (NA)      & NA (NA)       & NA (NA)       & NA (NA)     & NA (NA)       \\
\hline
\multicolumn{12}{c}{\textbf{Example 4 (Five-class Non-linear Example, MLG and MLDA are not Oracle)}}                                                                                                   \\ 
\hline
\textbf{EGKL} & \textbf{B1-SVM} & \textbf{BP1-SVM} & \textbf{B2-SVM} & \textbf{BP2-SVM} & \textbf{A-SVM} & \textbf{P-SVM} & \textbf{MLG} & \textbf{MLDA} & \textbf{TREE} & \textbf{RF} & \textbf{XGB}  \\ 
\hline
Time          & 1.0 (0.0)       & 1.0 (0.0)        & 0.8 (0.0)       & 0.8 (0.0)        & 7.2 (0.1)      & 2.1 (0.0)      & 0.0 (0.0)    & 0.0 (0.0)     & 0.3 (0.0)     & 1.5 (0.0)   & 0.5 (0.0)     \\ 
$L_1$            & 23.3 (0.3)      & 23.3 (0.3)       & 24.4 (0.7)      & 24.4 (0.7)       & 20.1 (0.2)     & 22.6 (0.2)     & 41.2 (0.1)   & 41.2 (0.1)    & 31.9 (0.2)    & 44.0 (0.1)  & 64.1 (0.2)    \\
$L_2$            & 5.3 (0.1)       & 5.3 (0.1)        & 5.6 (0.3)       & 5.6 (0.3)        & 4.6 (0.1)      & 5.2 (0.1)      & 7.1 (0.0)    & 7.2 (0.0)     & 6.1 (0.1)     & 9.9 (0.1)   & 21.6 (0.1)    \\
EGKL          & 12.0 (1.8)      & 12.0 (1.8)       & 12.5 (1.7)      & 12.5 (1.7)       & 10.1 (1.2)     & 11.9 (1.0)     & 18.0 (1.0)   & 18.6 (1.0)    & Inf (NaN)     & Inf (NaN)   & 65.1 (1.9)    \\
GKL           & 16.6 (0.7)      & 16.6 (0.7)       & 17.2 (1.0)      & 17.2 (1.0)       & 14.9 (0.5)     & 16.5 (0.4)     & 21.2 (0.2)   & 22.7 (0.3)    & Inf (NaN)     & Inf (NaN)   & 105.7 (1.2)   \\ 
TE1           & 41.5 (0.1)      & 41.4(0.1)        & 42.5 (0.4)      & 42.5 (0.4)       & 40.8 (0.1)     & 41.5 (0.1)     & 45.0 (0.1)   & 44.6 (0.1)    & 43.7 (0.2)    & 46.3 (0.1)  & 47.9 (0.1)    \\
TE2           & NA (NA)         & 41.3(0.1)        & NA (NA)         & 42.3 (0.4)       & NA (NA)        & 41.1 (0.1)     & NA (NA)      & NA (NA)       & NA (NA)       & NA (NA)     & NA (NA)       \\ 
$k^*$          & 5               & 5                & 1               & 1                & NA (NA)        & NA (NA)        & NA (NA)      & NA (NA)       & NA (NA)       & NA (NA)     & NA (NA)       \\ 
\hline
\textbf{GKL}  & \textbf{B1-SVM} & \textbf{BP1-SVM} & \textbf{B2-SVM} & \textbf{BP2-SVM} & \textbf{A-SVM} & \textbf{P-SVM} & \textbf{MLG} & \textbf{MLDA} & \textbf{TREE} & \textbf{RF} & \textbf{XGB}  \\ 
\hline
Time          & 1.0 (0.0)       & 1.0 (0.0)        & 0.8 (0.0)       & 0.8 (0.0)        & 7.2 (0.1)      & 2.1 (0.0)      & 0.0 (0.0)    & 0.0 (0.0)     & 0.3 (0.0)     & 1.5 (0.0)   & 0.5 (0.0)     \\ 
$L_1$            & 22.9 (0.3)      & 22.9 (0.3)       & 24.7 (0.7)      & 24.7 (0.7)       & 19.5 (0.2)     & 22.0 (0.2)     & 41.2 (0.1)   & 41.2 (0.1)    & 31.9 (0.2)    & 44.0 (0.1)  & 64.1 (0.2)    \\
$L_2$            & 5.2 (0.1)       & 5.2 (0.1)        & 5.5 (0.2)       & 5.5 (0.2)        & 4.5 (0.1)      & 5.0 (0.1)      & 7.1 (0.0)    & 7.2 (0.0)     & 6.1 (0.1)     & 9.9 (0.1)   & 21.6 (0.1)    \\
EGKL          & 11.8 (1.8)      & 11.8 (1.8)       & 12.3 (1.7)      & 12.3 (1.7)       & 9.8 (1.2)      & 11.1 (1.0)     & 18.0 (1.0)   & 18.6 (1.0)    & Inf (NaN)     & Inf (NaN)   & 65.1 (1.9)    \\
GKL           & 16.9 (0.6)      & 16.9 (0.6)       & 17.5 (0.9)      & 17.5 (0.9)       & 14.2 (0.4)     & 16.5 (0.3)     & 21.2 (0.2)   & 22.7 (0.3)    & Inf (NaN)     & Inf (NaN)   & 105.7 (1.2)   \\ 
TE1           & 41.5 (0.1)      & 41.4 (0.1)       & 42.6 (0.4)      & 42.6 (0.4)       & 40.9 (0.1)     & 41.4 (0.1)     & 45.0 (0.1)   & 44.6 (0.1)    & 43.7 (0.2)    & 46.3 (0.1)  & 47.9 (0.1)    \\
TE2           & NA (NA)         & 41.3 (0.1)       & NA (NA)         & 42.5 (0.4)       & NA (NA)        & 41.1 (0.1)     & NA (NA)      & NA (NA)       & NA (NA)       & NA (NA)     & NA (NA)       \\ 
$k^*$          & 5               & 5                & 1               & 1                & NA (NA)        & NA (NA)        & NA (NA)      & NA (NA)       & NA (NA)       & NA (NA)     & NA (NA)       \\
\hline
\end{tabular}}
\end{table}

 
\section{Benchmark Data Analysis}\label{sec:real-w}

We use five real data sets, including three low-dimensional cases and two high-dimensional cases, to compare the performance of all the methods. 

\subsection{Low-dimensional \orange{Cases} $(p<n)$}\label{reallowdim}
We first use three low-dimensional benchmark data sets, including pen-based handwritten digits, \emph{E.coli} and yeast protein localization data, to evaluate the wSVMs. The data information and performance measurements are shown in Table 3.

The pen-based handwritten digits data set, \orange{abbreviated as \emph{Pendigits}}, is obtained from \cite*{Dua_2019}. It consists of 7,494 training data points and 3,498 test data points, based on collecting 250 random digit samples from 44 writers \citep{alimoglu1996}. \orange{ Since the \emph{Pendigits} is a large data set with a total of $K=10$ classes, we start with a small classification problem by discriminating $K=4$ classes among digits 1, 3, 6, and 9; then we use the full data set to evaluate the methods for a $K=10$ class problem.}  The \emph{E.coli} data set is obtained from \cite*{Dua_2019}, which includes $n= 336$ samples and 
$p=7$ predictor variables. The goal is to predict the \emph{E.coli} cellular protein localization sites \citep{Horton1996}. Based on the biological functional similarity and locations proximity, we create a $K=4$ class problem, including class ``inner membrane'' with 116 observations, class ``outer membrane'' with 25 observations, class ``cytoplasm'' with 143 observations, and class ``periplasm'' with 52 observations. 
The yeast data set is obtained from \cite*{Dua_2019}, and the goal is to predict yeast cellular protein localization sites \citep{Horton1996} using 8 predictor variables. The total sample size is $n=1,484$ with 10 different localization sites. Based on the biological functional similarity and locations proximity, we formulate a $K=5$ class problem, including class ``membrane proteins'' with 258 observations, ``CYT'' with 463 observations, ``NUC'' with 429 observations,  ``MIT'' with 244 observations, and class ``other'' with 90 observations.

For the pen-based handwritten digits data set, we randomly split the training and tuning sets of equal sizes 10 times and report the average misclassification rate evaluated on the test set along with the standard error. For \emph{E.coli} and yeast data sets, since the class sizes are highly unbalanced, we first randomly split the data into the training, tuning, and test sets of equal size for each class separately, and then combine them across classes to form the final training, tuning, and test sets, for 10 times. We report the average test error rates along with the standard errors.

Table 3 suggests that the OVA-SVM overall is the best classifier among all cases, but its computational cost drastically increases with $n$ and $K$. The two linear time methods consistently give similar performance as pairwise coupling in terms of error rates, but with a significantly reduced computation time for large $n$ and $K$, and they outperform the benchmark methods. In some examples, the linear time method along with pairwise reconstruction may outperform pairwise \orange{coupling} and One-vs-All methods (e.g. BP1-SVM on yeast and Pendigits data sets). In practice, when $n$ and $K$ \orange{are} small, One-vs-All should be used for its best performance (e.g. \emph{E.coli} data set); when $K$ and $n$ \orange{are} large, linear methods have the best time complexity with similar performance as the pairwise and One-vs-All methods and are hence suitable for scalable applications.

\begin{table}[H]
\caption{Performance Measures for Benchmark Data Sets.}
\centering
\resizebox{\textwidth}{!}{
\begin{tabular}{c|ccc|c|cc|cc|c|c|ccccc} 
\hline

\multirow{2}{*}{\textbf{Data Sets}}                                          & \multicolumn{3}{c|}{\textbf{Data Info}}                          & \multirow{2}{*}{\textbf{Evals}} & \multicolumn{11}{c}{\textbf{Methods}}                                                                                                                                            \\ 
\cline{2-4}\cline{6-16}
                                                                            & \textbf{$\bm{K}$}       & \textbf{$\bm{n}$}            & \textbf{$\bm{\tilde{n}}$}        &                      & \textbf{B1-SVM} & \textbf{BP1-SVM} & \textbf{B2-SVM} & \textbf{BP2-SVM} & \textbf{A-SVM} & \textbf{P-SVM} & \textbf{MLG} & \textbf{MLDA} & \textbf{TREE} & \textbf{RF} & \textbf{XGB}  \\ 
\hline
\multirow{5}{*}{E.coli}                                                     & \multirow{5}{*}{4}  & \multirow{5}{*}{226}  & \multirow{5}{*}{110}  & Time                 & 0.1 (0.0)      & 0.1 (0.0)       & 0.1 (0.0)      & 0.1 (0.0)       & 0.3 (0.0)     & 0.2 (0.0)     & 0.0 (0.0)    & 0.0 (0.0)     & 0.0 (0.0)     & 0.1 (0.0)   & 1.3 (0.3)     \\ 
                                                                            &                         &                       &                       & TE1                  & 7.3 (0.5)      & 7.1 (0.4)       & 6.0 (0.4)      & 5.8 (0.3)       & 4.6 (0.4)     & 5.6 (0.3)     & 9.0 (1.3)    & 6.7 (0.7)     & 11.5 (0.7)    & 6.4 (0.6)   & 9.2 (1.2)     \\
                                                                            &                         &                       &                       & TE2                  & NA (NA)        & 6.9 (0.3)       & NA (NA)        & 5.6 (0.3)       & NA (NA)       & 5.5 (0.3)     & NA (NA)      & NA (NA)       & NA (NA)       & NA (NA)     & NA (NA)       \\ 

                                                                            &                         &                       &                       & $k^*$                    & 4              & 4               & 2              & 2               & NA~           & NA~           & NA           & NA~           & NA~           & NA~         & NA            \\ 
\hline
\multirow{5}{*}{Yeast}                                                      & \multirow{5}{*}{5}  & \multirow{5}{*}{990}  & \multirow{5}{*}{494}  & Time                 & 1.1 (0.0)      & 1.1 (0.0)       & 0.9 (0.1)      & 0.9 (0.1)       & 13.4 (0.0)    & 1.9 (0.0)     & 0.0 (0.0)    & 0.0 (0.0)     & 0.0 (0.0)     & 0.4 (0.0)   & 2.4 (0.2)     \\ 
                                                                            &                         &                       &                       & TE1                  & 39.3 (0.5)     & 38.1 (0.5)      & 41.7 (1.3)     & 41.5 (1.4)      & 39.0 (0.6)    & 39.4 (0.6)    & 40.0 (0.3)   & 40.5 (0.4)    & 43.0 (0.6)    & 37.9 (0.5)  & 42.4 (0.5)    \\
                                                                            &                         &                       &                       & TE2                  & NA (NA)        & 37.9 (0.6)      & NA (NA)        & 41.8 (1.4)      & NA (NA)       & 39.7 (0.4)    & NA (NA)      & NA (NA)       & NA (NA)       & NA (NA)     & NA (NA)       \\ 

                                                                            &                         &                       &                       & $k^*$                    & 1              & 1               & 2              & 2               & NA~           & NA~           & NA           & NA~           & NA~           & NA~         & NA            \\ 
\hline
\multirow{5}{*}{\begin{tabular}[c]{@{}c@{}}Pendigits\\ (1369)\end{tabular}} & \multirow{5}{*}{4}  & \multirow{5}{*}{2937} & \multirow{5}{*}{1372} & Time                 & 12.7 (0.1)     & 12.7 (0.1)      & 12.9 (0.3)     & 12.9 (0.3)      & 136.7 (0.9)   & 24.4 (0.3)    & 0.0 (0.0)    & 0.0 (0.0)     & 0.1 (0.0)     & 1.1 (0.0)   & 140.3 (4.7)   \\ 
                                                                            &                         &                       &                       & TE1                  & 1.0 (0.0)      & 0.9 (0.0)       & 1.0 (0.0)      & 1.0 (0.0)       & 0.8 (0.0)     & 0.9 (0.0)     & 3.3 (0.3)    & 6.8 (0.1)     & 7.1 (0.6)     & 1.3 (0.0)   & 2.0 (0.0)     \\
                                                                            &                         &                       &                       & TE2                  & NA (NA)        & 0.9 (0.0)       & NA (NA)        & 1.0 (0.0)       & NA (NA)       & 0.9 (0.0)     & NA (NA)      & NA (NA)       & NA (NA)       & NA (NA)     & NA (NA)       \\ 

                                                                            &                         &                       &                       & $k^*$                    & 1              & 1               & 1              & 1               & NA~           & NA~           & NA           & NA~           & NA~           & NA~         & NA            \\ 
\hline
\multirow{5}{*}{\begin{tabular}[c]{@{}c@{}}Pendigits\\ (Full)\end{tabular}} & \multirow{5}{*}{10} & \multirow{5}{*}{7494} & \multirow{5}{*}{3498} & Time                 & 43.4 (0.9)     & 43.5 (0.9)      & 53.1 (1.6)     & 53.1 (1.6)      & 1536.7 (9.4)  & 192.4 (1.9)   & 0.0 (0.0)    & 0.0 (0.0)     & 0.2 (0.0)     & 4.4 (0.1)   & 382.4 (5.4)   \\ 
                                                                            &                         &                       &                       & TE1                  & 4.1 (0.0)      & 3.9 (0.0)       & 4.5 (0.1)      & 4.4 (0.1)       & 3.7 (0.1)     & 4.0 (0.0)     & 10.6 (0.5)   & 17.1 (0.1)    & 25.9 (0.4)    & 5.3 (0.1)   & 5.2 (0.1)     \\
                                                                            &                         &                       &                       & TE2                  & NA (NA)        & 3.9 (0.0)       & NA (NA)        & 4.4 (0.1)       & NA (NA)       & 4.0 (0.0)     & NA (NA)      & NA (NA)       & NA (NA)       & NA (NA)     & NA (NA)       \\ 

                                                                            &                         &                       &                       & $k^*$                    & 2              & 2               & 1              & 1               & NA~           & NA~           & NA           & NA~           & NA~           & NA~         & NA            \\
\hline
\end{tabular}}
\end{table}

\noindent
\textbf{TABLE NOTE:} The first column gives the names of the \orange{four benchmark} data sets; the next 3 columns include the data set information: \orange{$K$ is the number of classes, $n$ is the training set size, and $\tilde n$ is the test set size.} The remaining columns compare the performance of the proposed methods with benchmarks.

\medskip

The results suggest that the proposed One-vs-All method (OVA-SVM) consistently outperforms all of the methods in all the data sets, but has a high computational cost; the proposed baseline learning methods (B-SVM and BP-SVM) have a close performance to the pairwise coupling method and consistently outperform the benchmark methods, with a significant reduction in \orange{computational} time. Similar observations are made for these four studies.

\subsection{High-Dimensional Cases $(p > n)$}\label{highdimexp}
Two high-dimensional data sets are downloaded from the site:  \url{https://schlieplab.org/Static/Supplements/CompCancer/datasets.htm}. The first data set consists of microarray gene expressions for the three major adult lymphoid malignancies of B-cell lymphoma: Diffuse large B-cell lymphoma (DLBCL), follicular lymphoma (FL), and chronic lymphocytic leukemia (CLL), as discussed in \cite{alizadeh2000}. There are a total of 2,093 expression profiles from 62 subjects. The second data set is the oligonucleotide microarrays gene expression profile for the treatment of pediatric acute lymphoblastic leukemia (ALL) based on the patient's risk of relapse, with six prognostically leukemia subtypes, including T-ALL, E2A-PBX1, BCR-ABL, TEL-AML1, MLL rearrangement, and hyperdiploid $>50$ chromosomes. We combine the TEL-AML1, BCR-ABL and hyperdiploid $>50$ chromosomes as class T+B+H for their similar gene functional clustering as discussed in \cite{yeoh2002} and make it a $K=4$ class problem, with $n=248$ and $p=2,526$. 
For each class, we randomly sample 75\% of the data points and then spit them equally into the training and tuning sets, and the remaining 25\% of the data points are used as the test set. These three sets are combined across all the classes to form the final training, tuning, and test sets. For the B-cell lymphoma data set, we have the training and tuning sets of size 48 (containing 32 DLBCL, 7 FL, 9 CLL) and the test set of size 14 (containing 10 DLBCL, 2 FL, 2 CLL). For the acute lymphoblastic leukemia data set, we have the training and tuning set of size 188 (containing 33 T-ALL, 119 T+B+H, 15 MLL, 21 E2A-PBX1) and the test set of size 60 (containing 10 T-ALL, 39 T+B+H, 5 MLL, 6 E2A-PBX1).

\begin{figure}[H]
\centering
\caption{This figure plots the estimated probabilities (shown as the bar heights) for the B-cell lymphoma data set's test set for 14 data points with 3 methods. From the top to the bottom row are the evaluation of the pairwise coupling (P-SVM), \orange{OVA-SVM} (A-SVM), and baseline (B1-SVM) methods. From the left to the right columns are estimated class probabilities of one subclass of test samples: DLBCL, CLL, and FL, labeled as 1,2,3, respectively, by the three methods. Three colors (red, azure, green) represent the estimated probabilities of test data points belonging to class DLBCL, CLL, and FL.}

    
        \includegraphics[max size={\textwidth}{\textheight}]{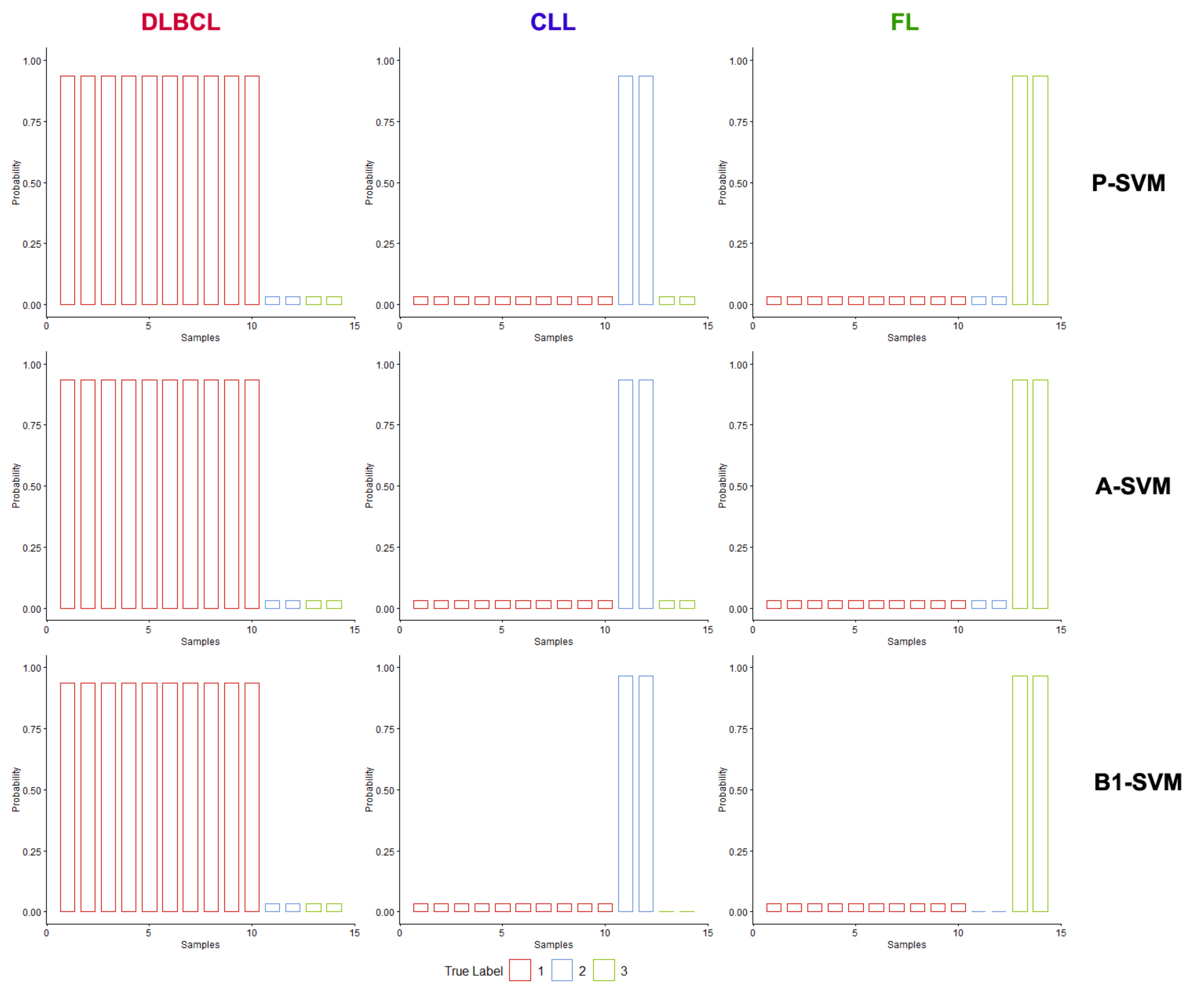}
              \phantomsubcaption
        \label{fig:1f}

\end{figure}

\begin{figure}[H]
\centering
\caption{This figure plots the estimated probabilities (shown as the bar heights) for the acute lymphoblastic leukemia test set for 60 data points with 3 methods. From the top to the bottom row are the pairwise coupling (P-SVM), \orange{OVA-SVM} (A-SVM), and the baseline (B2-SVM) methods. From the left to the right are estimated class probabilities of one subclass of test samples: T-ALL, T+B+H, MLL, and E2A-PBX1, labeled as 1,2,3,4 respectively, by the three methods. Four colors (blue, orange, green, garnet) represent the estimated probabilities of test data belonging to class T-ALL, T+B+H, MLL, and E2A-PBX1.}

        
          \includegraphics[max size={1\textwidth}{\textheight}]{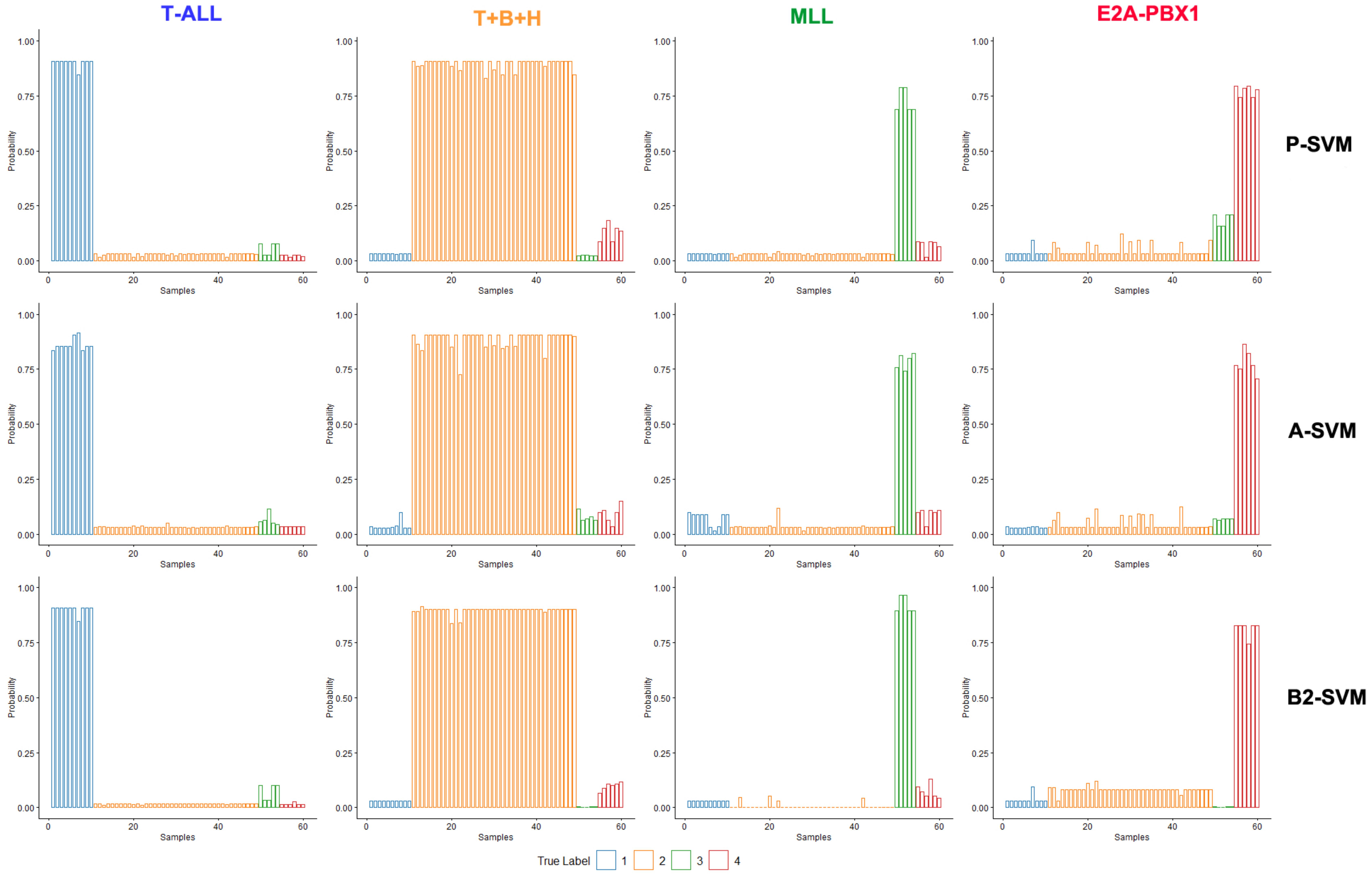}
              \phantomsubcaption
        \label{fig:2f}
\end{figure}

For both data sets, we normalize the gene expression values, obtain their $Z$-scores, and rank the genes by their BW criterion of the marginal relevance to the class label, calculated as the ratio of each gene's between-group to within group sums of squares, as specified in \cite{Dudoit02}. We select top 5 genes as predictor variables based on the order of the relevance measures of the BW score. Figure 2 plots the estimated class probability for B-cell lymphoma data set on 14 test data points with the baseline method 1 (B1-SVM), \orange{OVA-SVM} (A-SVM), and pairwise coupling (P-SVM), whose \orange{running} time of 1 sec, 3 secs, and 2 secs, respectively. 

Figure 3 plots the estimated class probability for acute lymphoblastic leukemia data set on 60 test data points with baseline method 2 (B2-SVM), \orange{OVA-SVM} (A-SVM), and pairwise coupling (P-SVM), whose \orange{running} time of 5 secs, 18 secs, and 13 secs. For both data sets, the two baseline methods and OVA-SVM have similar class probability estimations for each test data point as pairwise coupling, with no misclassification for any classes (we illustrated one baseline model for each data set). However, the baseline methods have the fastest \orange{computational} time. With the appropriate variable selection method, our proposed linear algorithms and One-vs-All methods can work with scalable high-dimensional data with high accuracy on multiclass probability estimation and classification.

\section{Concluding Remarks}\label{sec:cond}

In this article, we propose two new learning schemes, the baseline learning and the \orange{One-vs-All} learning, to improve the computation time and estimation accuracy of the wSVMs methods. The One-vs-All method shows consistent best performance in probability estimation and classification with the most expensive computational cost, which is good for a small sample size with a class size $K \le 5$. The two baseline methods have comparable performance \orange{to} pairwise coupling method with great value in reducing the time complexity especially when class size $K$ and sample size $n$ is big, which makes it a favorable approach for multiclass probability estimation in real-world applications for both low-dimensional and high-dimensional data analysis. 

We also establish the consistency theory for the baseline models and One-vs-All multiclass probability estimators, which is similar \orange{to the} pairwise coupling method. The nature of the baseline method for training $K-1$ binary wSVMs, makes it enjoys the advantage of parallel computing and \orange{adapts the general-purpose computing on graphics processing units (GPGPU)}, which could further reduce the \orange{computational time} and \orange{optimized} for complex multiclass probability estimation problems. One interesting question is how to use our proposed \orange{wSVMs} framework for image classification, such as brain tumor MRI images, for fast screening. Since the widely used image classification approaches based on Convolutional Neural Networks (CNNs) \orange{lack the} accurate class probability estimation \citep{guo2017,minderer_revisiting_2021}, which is important for pre-cancer diagnosis to establish the strength of confidence for classification.

Currently, we only evaluate the dense feature set. The class probability estimation performance would decrease significantly if the feature sets are sparse with highly correlated predictor variables and abundant noisy features (as shown in the high-dimensional data sets in \ref{highdimexp}). Since all the methods discussed in this article are based on the $L_2$-norm regularization of the binary wSVMs optimization problem, we don't have the automatic variable selection power with the current framework. One of the important directions for future research is how to incorporate variable selection in the proposed class probability estimation framework for wSVMs, which is an important problem for data with sparse signals and highly correlated features. In many real applications, abundant features are collected but only \orange{a} few of them \orange{have} predictive power. We will this topic a more thorough treatment in future work. 

\medskip

\section*{Supplementary Material} 
\orange{We developed the R Package \texttt{MPEwSVMs} for multiclass probability estimation based on the proposed methods and have submitted it to the GitHub repository at \url{https://github.com/zly555/MPEwSVMs}. Detailed instructions on how to use the package and as well as a numerical simulation example are shown in the \texttt{README.md} file.} 

\medskip
\bibliography{reflist}
\bibliographystyle{ims}

\end{document}